\documentclass[pra,twocolumn,showpacs,tightenlines,
preprintnumbers,amsmath,amssymb]{revtex4}
\usepackage{graphicx}
\usepackage{epstopdf}
\newcommand{\beq}{\begin{equation}}
\newcommand{\eeq}{\end{equation}}
\newcommand{\bea}{\begin{eqnarray}}
\newcommand{\eea}{\end{eqnarray}}
\newcommand{\ba}{\begin{array}}
\newcommand{\ea}{\end{array}}
\newcommand{\bc}{\begin{center}}
\newcommand{\ec}{\end{center}}

\newcommand{\bml}{\begin{mathletters}}
\newcommand{\eml}{\end{mathletters}}
\newcommand{\commentout}[1]{{}}
\newcommand{\bk}{{\bf k}}

\newcommand{\q}{{\bf q}}
\newcommand{\p}{{\bf p}}
\newcommand{\br}{{\bf r}}

\newcommand{\half}{\hbox{$1\over2$}}

\newcommand{\eq}[1]{(\ref{#1})}

\newcommand{\comment}[1]{{}}

\begin{document}
\title{Mean-field stationary state of a Bose gas at a Feshbach
resonance}
\author{Andrew Carmichael}
\author{Juha Javanainen}
\affiliation{Department of Physics, University of Connecticut,
Storrs, CT 06269-3046}
\date{\today}
\pacs{03.75.Hh,03.75.Ss,03.75.Lm}

\begin{abstract}
We study the steady state of a zero-temperature Bose gas near a Feshbach
or photoassociation resonance using a two-channel mean-field model
that incorporates atomic and molecular condensates, as well as correlated atom
pairs originating from dissociation of molecules into pairs of atoms. We
start from a many-body Hamiltonian for atom-molecule conversion, and derive
the time dependent version of the mean-field theory. The stationary solution
of the time dependent model is rendered unique with an approximation
that entails that all noncondensate atoms are
correlated, as if emerging from dissociation of molecules. The steady state
is solved numerically, but limiting
cases are also found analytically. The system has a phase transition in which the atomic
condensate emerges in a nonanalytic fashion. We quantify the scaling of
the observable quantities, such as fractions of atomic and molecular
condensates, with the detuning and the atom-molecule
conversion strength. Qualitatively, the dependence on detuning rounds out
with increasing coupling strength. A study of the thermodynamics shows that
the pressure of the atom-molecule system is negative,
even on the molecule side of the resonance. This indicates the possibility of
mechanical instability.
\end{abstract}
\pacs{}
\maketitle

\section{Introduction}
Photoassociation~\cite{THO87,NAP94} and especially the mathematically
equivalent Feshbach resonance~\cite{STW76,TIE92} have been main themes in the
physics of quantum degenerate Bose and Fermi gases for a while. The
earliest experimental work was carried out on bosons. The widest-known
examples are probably Feshbach resonance in a BEC~\cite{INO98}, two-photon
photoassociation~\cite{WYN00} in a BEC, purported~\cite{WUS07} mechanical
collapse of a BEC when the atom-atom interaction is made attractive by using
a Feshbach resonance~\cite{DON01}, and Ramsey fringes in the conversion
between atomic and molecular condensates~\cite{DON02}.

In an aspect that is often concealed by phrases to the effect that the
Feshbach resonance is used to tune atom-atom interactions, at least in
principle photoassociation and Feshbach resonance always involve conversion
of atom pairs to the corresponding diatomic molecules. These molecules are in
a highly excited vibrational state and are prone to collisional quenching. In
a Bose gas the molecules tend to be  short-lived ($\sim$ 10 ms). However, it
turns out~\cite{STR03,PET04a} that the diatomic molecules created at the
$834\,{\rm G}$  Feshbach resonance
from fermionic ${}^6$Li atoms may persist for seconds.
Thermal-equilibrium experiments are feasible, which has spurred an
enormous experimental and theoretical interest. Molecular condensates are now
prepared routinely~\cite{JOC03,ZWI03}. Of particular interest to our theme of
atom-molecule conversion is a measurement of the equilibrium fraction of
molecules as a function of the magnetic field~\cite{PAR05}, while a comparison
of the thermodynamical properties between experiments~\cite{LUO07} and
quantum Monte Carlo simulations~\cite{BUL07} serves as an illustration of the
present state of the art in the modeling of fermion systems.

Given that the initial Feshbach resonance experiments were time dependent in
an essential way, the corresponding theories were time dependent as well.
Our approach is based on field-theoretical modeling of atom-molecule
systems~\cite{TIM98,DRU98,JAV99}, but there are methods springing, e.g.,
from theory of molecular structure, as reviewed recently in
Ref.~\cite{KOH06}. For instance, a good agreement with the
atom-molecule Ramsey fringe experiments  in a BEC~\cite{DON02} has been
reported by several groups~\cite{KOK02a,MAC02,KOH03,DUI03,MAC02y}. Our time
dependent method has produced~\cite{JAV04} a passable
theoretical description of atom-molecule conversion in an
experiment~\cite{REG03}, and yielded valid~\cite{HOD05} predictions about
the temperature dependence and the maximum value of the conversion
efficiency also for fermionic atoms.

With the emergence of  thermal-equilibrium experiments in Fermi systems, the equilibrium-oriented theoretical machinery of condensed matter physics has been brought to bear~\cite{CAR03,CHE05a,BUL07}. On the other hand, we
have noticed~\cite{JAV05} that the stationary solution of our time dependent
formalism for fermions~\cite{JAV04} may also serve as a zero-temperature
thermal equilibrium theory. The stationary solution is not unique, but it may
be made so by assuming that all fermions appear in correlated pairs, as if
from dissociation of molecules. The result turns out to be a variant of the
atom-molecule version of the BCS theory~\cite{EAG69,LEG80,NOZ85,RAN95} for
Fermi gases. We~\cite{JAV05} and others~\cite{CHE05a,ROM05} have reported
favorable theoretical comparisons with the equilibrium fraction of molecules in the
experiments~\cite{PAR05} in which the magnetic field was varied in the
neighborhood of the broad ${}^6$Li Feshbach resonance.

The immediate purpose of the present paper is to introduce a similar ``BCS''
theory for bosonic atoms at zero-temperature thermal equilibrium. Basically,
we take our time dependent theory~\cite{JAV02,MAC02,MAC02y} and
find the time-independent solution. As with fermions, the stationary solution
may be rendered unique with a pairing approximation, which, however, turns
out to be more subtle for bosons than for fermions. As the secondary goal, we
discuss the technicalities of our work on both
bosons~\cite{JAV02,MAC02,MAC02y} {\em and\/} fermions~\cite{JAV04,JAV05} that
were not detailed in the original letter format publications.

The main qualitative finding is that the atom-molecule system exhibits a phase
transition in which the atomic BEC emerges in a nonanalytic fashion when the detuning,
the atom-molecule energy difference, is varied. In the limit of weak
atom-molecule conversion, for instance in the limit of a very dilute gas, the
phase transition is at the position of the two-body Feshbach resonance, but for increasing
atom-molecule conversion strength it moves to the molecule
side of the Feshbach resonance or photoassociation. We
characterize the phase transition and the variation of quantities
such as the molecule fraction with both the detuning and the atom-molecule
conversion strength. Broadly speaking, the dependence on detuning rounds
out with increasing interaction strength. Finally, we investigate the
thermodynamics of the atom-molecule system. Our present model with
a Feshbach or photoassociation resonance but no attendant background
scattering length produces a puzzling surprise: The calculated
pressure of the gas is negative for all parameter values, which indicates
mechanical instability.

\section{Formulation of the problem}
\subsection{Hamiltonian}
We model conversion of bosonic atoms into bosonic diatomic molecules
using the momentum (wavevector) representation Hamiltonian~\cite{JAV99,KOS00}
\bea
\frac{H}{\hbar} &=& \sum_{\bk} \left[
\epsilon_\bk\, a^\dagger_\bk a_\bk + (\delta+\half\epsilon_\bk)
b^\dagger_\bk b_\bk
\right]\nonumber\\
&&-\half\,\sum_{\p,\q} (\kappa_{\p,\q}\,b^\dagger_{\p+\q}a_\p a_\q +
\kappa_{\p,\q}^*\,a^\dagger_\q a^\dagger_\p b_{\p+\q})\,,
\label{HAM}
\eea
as appropriate for a free (non-trapped) gas.
Here $a_\bk$ and $b_\bk$ are boson annihilation operators for atoms and
molecules with momentum $\hbar\bk$, and $\hbar\epsilon_\bk \equiv
\hbar^2\bk^2/2m$ is the kinetic energy for an atom with wave vector
$\bk$. For a molecule with twice the mass of an atom, the kinetic energy for a
given momentum is half of the energy of an atom. The
detuning $\delta$ gives the energy difference between a stationary
molecule ($\bk=0$) and two stationary atoms in the form $\hbar\delta$.
This parameter is varied in a Feshbach resonance by varying the magnetic
field, and in photoassociation by tuning the frequency
(frequencies) of the laser(s).

Atom-molecule conversion entails that a pair of atoms is either
converted to a molecule or a molecule is dissociated to a pair of atom, all
the while conserving the momentum. The governing matrix elements are denoted
by
$\kappa_{\p,\q}$. Ordinarily we deal with $s$-wave processes that dominate at
low temperature. By virtue of translational and rotational invariance, the
corresponding coupling matrix element is then of the form
$\kappa_{\p,\q}=\kappa(|\p-\q|)$. Furthermore, we mostly write the coupling
matrix element as a once-and-for all constant $\kappa_{\p,\q}=\kappa$. This
means that atom-molecule conversion is a zero-range contact
interaction; in the position representation for the atomic and molecular fields
$\hat\phi$ and $\hat\psi$, the corresponding term in the Hamiltonian
density would be
$\propto\kappa[{\hat\psi}^\dagger(\br){\hat\phi}(\br){\hat\phi}(\br)+
{\hat\phi}^\dagger(\br){\hat\phi}^\dagger(\br)
{\hat\psi}(\br)]$.  The contact interaction model is convenient for our
aims for two reasons: it reduces the number of parameters to
consider, and endowing the coupling coefficients with a realistic momentum
dependence would much complicate the solution of the model. The downside is an
ultraviolet divergence~\cite{JAV02}, which will require a renormalization.

Consider two atoms that may combine into a molecule in the
center-of-mass frame. Then a molecular bound state is coupled to the
continuum of the relative motion of the two atoms. By redefining the global phases
of the continuum wave functions, one may always arrange things so that the coupling
coefficients 
$\kappa_{\p,\q}$ and $\kappa$ are real, and $\kappa$ is also non-negative.
This is assumed to be the case below. 
We write {\em sums\/} over the wave vectors in
Eq.~\eq{HAM}, which presumes box normalization of the dissociated states. The
quantization volume
$V$ therefore enters the matrix element, $\kappa_{\p,\q}\propto
V^{-1/2}$~\cite{JAV99,KOS00}.

Conspicuously missing from the model
is a background scattering length that would prevail, say, far away
from the Feshbach resonance. It could, and given the nature of our results,
maybe should be included. But again, since our focus is on the nature of
the system in the vicinity of the resonance $\delta=0$, we keep the model
and the attendant technical complications to the bare minimum.

A break-up of a molecule produces two atoms, and it takes precisely two
atoms to make a molecule. The Hamiltonian~\eq{HAM} correspondingly has
the conserved quantity
\beq
\hat{N} = \sum_\bk(a^\dagger_\bk a_\bk + 2 b^\dagger_\bk b_\bk)\,,
\label{INVNUM}
\end{equation}
number of atoms plus twice the number of molecules. We use the value $N$ of
the invariant $\hat N$ to characterize the number of particles in the
system, and occasionally, slightly inaccurately, call it the atom number.
The total momentum
\begin{equation}
\hat{\bf P} = \hbar\sum_\bk \bk(a^\dagger_\bk a_\bk+b^\dagger_\bk b_\bk)
\end{equation}
is likewise a constant of the motion.
\subsection{The dressed molecule}
\label{TWOATOMPROBLEM}
Before proceeding to many-body systems, we investigate the solvable
model with the invariant atom number equal to two. Without restricting
the generality, we also assume that the conserved center-of-mass momentum
equals zero. The state space is then spanned by vectors of the
form~\cite{JAV05}
\begin{equation}
|\psi\rangle =\left( \sum_\bk A(\bk) a^\dagger_\bk
a^\dagger_{-\bk}+\beta b^\dagger_0\right)|0\rangle\,,
\end{equation}
where $A(\bk)$ [with $A(\bk)=A(-\bk)$] and $\beta$ are complex numbers,
and
$|0\rangle$ is the particle vacuum. The time dependent Schr\"odinger
equation follows from Hamiltonian~\eq{HAM} in the form
\bea
i\dot\beta &=& \delta\beta-\frac{\kappa}{2}\sum_\bk A({\bk})\,,\\
i \dot{A}(\bk) &=&2\epsilon_\bk A(\bk) - \kappa\beta\,.
\eea

If rotational invariance prevails at the initial time so that $A(\bk)$ is
only a function of $|\bk|$, or equivalently, a function of
$\epsilon_\bk\equiv\epsilon$, the same symmetry holds at all later times.
Moreover, in a two-atom system we may replace the sum over $\bk$ by a
continuum approximation without running into problems with the atomic BEC. We
write the continuum approximation as
\begin{equation}
\sum_\bk f(\epsilon(\bk))\!\rightarrow\!\frac{V}{(2\pi)^3}\int d^3k \,
f(\epsilon(\bk))
\!\rightarrow\! \frac{3N}{2\epsilon_F^{3/2}}\int_0^\infty
d\epsilon\,\sqrt\epsilon f(\epsilon)\,.
\label{CONTAPP}
\end{equation}
We prefer quantities with the dimension of frequency over
$\hbar$ times the same quantities with the dimension of energy, so that
the integral runs over frequencies. Although it is silly if not
misleading in the present case of only two atoms ($N=2$), for future use we
have defined  the energy $\hbar\epsilon_F$ by
\begin{equation}
\epsilon_F=\frac{\hbar}{2m}\left(\frac{6\pi N}{V}\right)^{2/3}\,.
\end{equation}
It equals the Fermi energy for a single-component gas with the density
$N/V$, but obviously has nothing to do with any physical Fermi energy.
Instead,
$\epsilon_F$ is a measure of the density of the gas; the
essentially unique frequency that can be construct out of density
($N/V$) for a quantum mechanical ($\hbar$) gas of atoms ($m$). Below we will refer to $\epsilon_F$ {\em without\/} the $\hbar$ as Fermi energy.

To cap our introduction of the notations, we define the frequency parameter
characterizing the atom-molecule coupling
\begin{equation}
\Omega = \sqrt N \,\kappa\,.
\end{equation}
It is essential to keep in mind that in this bound-continuum problem
the analogy of the Rabi frequency $\Omega$ is proportional to the square root
of density, $\Omega\propto(N/V)^{1/2}$. In line of what was said before
of the coefficient $\kappa$, we also take $\Omega \ge 0$. The time dependent
Schr\"odinger equation for the coefficients $\beta$ and $A(\epsilon(\bk))$
finally reads
\bea
i\dot\beta(t)&=&\delta\beta(t)-\frac{3\Omega}{2\sqrt{2}\epsilon_F^{3/2}}\int_0^\infty
d\epsilon\,\sqrt\epsilon\, A(\epsilon,t)\,,\label{SCHRBETAEQ}
\\
 i\dot
A(\epsilon,t)&=&2\epsilon
A(\epsilon,t)-{\Omega\over\sqrt2}\beta(t)\,.\label{SCHRAEQ}
\eea

The essence of the two-channel theory is to regard atoms and molecules as
distinct though coupled degrees of freedom. The boson operators in
Hamiltonian~\eq{HAM} create and annihilate bare atoms and molecules that would
be observed if there were no atom-molecule coupling. As such, they would
represent the observable atoms and molecules immediate after the atom-molecule
coupling were switched off. In the case of photoassociation this could be
achieved literally by switching off the lasers. For the Feshbach
resonance an equivalent decomposition could be effected (in principle) by
suddenly switching the magnetic field so far off the resonance that the atoms and
the molecules effectively decouple. There are also experimental probes that
directly see the bare molecules, for instance, by making use of optical
transitions in the bare molecules~\cite{PAR05}.

However, standard radio frequency spectroscopy at a Feshbach
resonance~\cite{REG03,BAR05} probes transitions between energy eigenstates of
the system in the presence of the atom-molecule coupling, i.e., stationary
states of Eqs.~\eq{SCHRBETAEQ} and~\eq{SCHRAEQ}. These are superpositions of a
bare molecule and pairs of bare atoms, and so we refer to the coupled system
as the dressed molecule.

\subsubsection{Renormalization}
The energy eigenstates are obtained by inserting an Ansatz of the form
$\beta(t) =e^{-i\omega t}\beta$, $A(\epsilon,t) =e^{-i\omega t}A(\epsilon)$ into
Eqs.~\eq{SCHRBETAEQ} and~\eq{SCHRAEQ}, which gives
\bea
(\omega-\delta)\beta&=&-\frac{3\Omega}{2\sqrt{2}\epsilon_F^{3/2}}\int_0^\infty
d\epsilon\,\sqrt\epsilon\, A(\epsilon)\,,\label{STATBETAEQ}\\
 \dot
(\omega-2\epsilon)A(\epsilon)&=&-{\Omega\over\sqrt2}\beta\,.\label{STATAEQ}
\eea
Simple elimination of $A(\epsilon)$ from Eq.~\eq{STATBETAEQ} using
Eq.~\eq{STATAEQ} gives a relation to determine the eigenfrequency
$\omega$,
\begin{equation}
\omega-\delta = \frac{3\Omega^2}{4\epsilon_F^{3/2}}\int_0^\infty
d\epsilon\frac{\sqrt{\epsilon}}{\omega-i\eta-2\epsilon}\,.
\label{TWOPARTEVE}
\end{equation}
Here $-\eta$, with $\eta=0+$, is the usual imaginary part in the energy that
needs to be added to handle the divergence of the integrand at
$2\epsilon=\omega$. This practice is the same as if we took Fourier
transformations of the time dependent equations and used them to study the
evolution of the system forward in time. Such an asymmetry
in the direction of time is not desirable if we are looking for true
stationary states of the atom-molecule system. We outline in
Appendix~\ref{FANOSS} a method, following Ref.~\cite{FAN61}, to find the proper
stationary states, but here the main issue is the ultraviolet divergence of the integral in~\eq{TWOPARTEVE}.

Physically, Eqs.~\eq{SCHRBETAEQ} and~\eq{SCHRAEQ} describe the coupling of the
bare-molecule state to many (actually, a continuum  of) atom-pair states. One
obvious consequence is that, if there still is a bound state in the system, it is shifted in energy from the original
bare molecular state. In the contact-interaction model the shift simply is infinite.

We renormalize as follows~\cite{JAV02,JAV05}. We adopt an upper limit of the
integral
$M$, write Eq.~\eq{TWOPARTEVE} as 
\bea
\omega&-&\left(\delta-\frac{3\Omega^2}{4\epsilon_F^{3/2}}\int_0^M
\frac{\sqrt{\epsilon}}{2\epsilon}
\right)\nonumber\\
& &= \frac{3\Omega^2}{4\epsilon_F^{3/2}}\int_0^M
d\epsilon\left(\frac{\sqrt{\epsilon}}{\omega-i\eta-2\epsilon}
+\frac{\sqrt{\epsilon}}{2\epsilon}\right)\,,
\label{UPLIM}
\eea
and let $M\rightarrow\infty$ at the end of the calculation. The right-hand
side then converges nicely, but ostensibly not so the left-hand side; the
detuning
$\delta$ gets modified by the infinite level shift. The idea of the renormalization is to
incorporate the level shift into the definition of the energies, and take the
renormalized detuning
\begin{equation}
\lim_{M\rightarrow\infty}
\,\,\delta-\frac{3\Omega^2}{8\epsilon_F^{3/2}}\int_0^M\frac{1}{\sqrt{\epsilon}}
= \bar\delta
\end{equation}
to have a finite value.

In this way Eq.~\eq{TWOPARTEVE} turns into a well-behaved equation
\begin{equation}
\omega-\bar\delta = \frac{3\Omega^2}{4\epsilon_F^{3/2}}\int_0^\infty
d\epsilon\,\frac{1}{\sqrt{\epsilon}}\,\frac{\omega}{\omega+i\eta-2\epsilon}
\,.
\label{RENODEF}
\end{equation}
The salient point is that for $\bar\delta<0$ Eq.~\eq{RENODEF} has precisely one
real solution (with $\omega<0)$, and no real solution if $\bar\delta>0$. The
system has a true
stationary state that does not evolve in time only for $\bar\delta<0$. In
Appendix~\ref{FANOSS} we replace this statement with the more precise
observation that the dressed molecule has a {\em normalizable\/} stationary state if
and only if $\bar\delta<0$. A bound state is found for the dressed molecule
for $\bar\delta<0$, otherwise the dressed molecule only exists in a
dissociated form as a pair of bare atoms with a component of the bare
molecule mixed in. We take this to mean that the renormalized detuning
$\bar\delta=0$ denotes the position of the Feshbach resonance in the
two-atom system.

\subsection{Mean-field approximation}
We make the many-particle system solvable by a mean-field approximation, the idea
of which is to treat possible atomic and molecular condensates as classical
fields not quantum fields anymore~\cite{TIM98,JAV99,JAV04}. Although it is not
essential for the structure of the mean-field theory but rather a technical
assumption to facilitate the analysis, we also assume that all molecules
present in the system belong to a condensate of zero-momentum molecules. Hence,
only the molecule operator $b_0\equiv b$ is kept in the Hamiltonian. Moreover,
to accommodate the corresponding atomic BEC, we already at this point track
separately the zero-momentum atoms with $a_0\equiv a$. Using the Hamiltonian~\eq{HAM}
we then find the Heisenberg picture equations of motion for the atomic and molecular operators
\bea
i\dot a &=& -\kappa ba^\dagger,\label{FHEQ}\\
i\dot b &=& \delta b -\half\kappa aa-\half\kappa\sum_\bk \, a_\bk
a_{-\bk}\,,
\\ i\dot a_\bk &=& \epsilon_\bk a_\bk-\kappa b a^\dagger_{-\bk}\,.
\eea
Out of these primary equations one may form equations of motion for
quadratic operator products, e.g.,
\bea
i\frac{d}{dt}(a^\dagger_{\bk}a_{\bk})&=&\kappa(b^\dagger a_{\bk}a_{-\bk}-
a^\dagger_{-\bk}a^\dagger_{\bk}b)\,,
\\
i\frac{d}{dt}(a_{\bk}a_{-\bk}) &=& 2\epsilon_\bk
\,a_{\bk}a_{-\bk}\nonumber\\
&&-\kappa (1+a^\dagger_{\bk}a_\bk +
a^\dagger_{-\bk}a_{-\bk})b\,.
\label{LHEQ}
\eea

We implement the mean-field approximation at this juncture by stating
that in the equations of motion $a$ and $b$ are $c$-numbers not operators
anymore~\cite{JAV02,JAV04}. The quantum mechanical expectation values for the products then factorize
as in this example,
\begin{equation}
\langle a^\dagger_\bk a_\bk b \rangle = \langle a^\dagger_\bk
a_\bk\rangle   b \equiv  \langle a^\dagger_\bk
a_\bk\rangle \langle b \rangle\,.
\end{equation}
Applying the factorization to Eqs.~\eq{FHEQ}--\eq{LHEQ}  gives a 
closed set of equations of motion involving the expectation values of
the form
$\langle a\rangle$,
$\langle b\rangle$,
$\langle a^\dagger_\bk a_\bk\rangle$, and $\langle a_\bk
a_{-\bk}\rangle$.

For the convenience of the formulation we assume that the problem
is effectively rotationally symmetric, so that, e.g., the expectation
value $\langle a_\bk a_{-\bk}\rangle$ only depends on the energy
$\epsilon\equiv\epsilon_\bk$. We write
\begin{equation}
P(\epsilon) = \langle a^\dagger_\bk a_\bk\rangle,\quad
A(\epsilon) = \langle a_\bk a_{-\bk}\rangle\,.
\end{equation}
It should be noted that $P(\epsilon)$ stands for the
expectation value of the number of atoms in a one-particle state with
the energy $\hbar\epsilon$, not for a quantity such as the number of atoms per
unit energy interval. We also define the amplitudes for atomic and
molecular condensates
\begin{equation}
\alpha = \sqrt\frac{1}{N}\, a,\quad \beta = \sqrt\frac{2}{N}\, b
\end{equation}
so that $|\alpha|^2$ and $|\beta|^2$ stand for the fractions of the atoms
that are in the system as part of either the atomic or the molecular
condensate. Finally, as the atomic BEC has already been taken into account
separately, the continuum approximation~\eq{CONTAPP} should work as before.

We finally have the equations of
motion of our mean-field theory for atom-molecule conversion in a boson
system,
\bea
i\dot\alpha(t) &=&
-{\Omega\over\sqrt{2}}\,\beta(t)\alpha^*(t),\label{ALPHAEQ}\\
i\dot\beta(t)&=&\delta\beta(t)-{\Omega\over\sqrt{2}}\,\alpha^2(t)\nonumber\\
&&-\frac{3\Omega}{2\sqrt{2}\epsilon_F^{3/2}}\int
d\epsilon\,\sqrt\epsilon\, A(\epsilon,t),\label{BETAEQ}
\\
 i\dot
A(\epsilon,t)&=&2\epsilon
A(\epsilon,t)-{\Omega\over\sqrt2}\,[1+2P(\epsilon,t)]\beta(t),\label{AEQ}\\
i\dot P(\epsilon,t)&=&{\Omega\over\sqrt{2}}[\beta^*(t)A(\epsilon,t)-
\beta(t) A^*(\epsilon,t)]\label{PEQ}\,.
\eea

The similarity to the notation we employed in the discussion of the two-atom
problem in Sec.~\ref{TWOATOMPROBLEM} is no accident. In fact, in the absence
of a BEC of atoms, and assuming that the occupation numbers of the atomic
states $P(\epsilon,t)$ are negligible compared to unity, Eqs.~\eq{ALPHAEQ} and~\eq{BETAEQ}
coincide with the two-atom theory Eqs.~\eq{SCHRBETAEQ} and~\eq{SCHRAEQ}. We
view the mean-field theory as the two-atom theory amended with the possibility
of a BEC and Bose enhancement for the atoms. An analogous interpretation applies to the
corresponding mean-field theory for fermions	\cite{JAV05}.

Writing the expectation value of the invariant atom number~\eq{INVNUM}
in terms of the mean-field variables gives the equation
\begin{equation}
|\alpha|^2+|\beta|^2 + \frac{3}{2\epsilon_F^{3/2}}\int
d\epsilon\,\sqrt\epsilon\, P(\epsilon)=1\,.\label{NORMEQ}
\end{equation}
The left-hand side is indeed a constant of the motion under
Eqs.~\eq{ALPHAEQ}--\eq{PEQ}, so that the present mean-field approximation
successfully reflects a basic property of the Hamiltonian.  Finally, one may
write the expectation value of the Hamiltonian in the mean-field
approximation as
\bea
e&=&\frac{\langle H \rangle}{\hbar N}\nonumber\\
&=&
\half\delta|\beta|^2+\frac{3}{2\epsilon_F^{3/2}}\int
d\epsilon\,\epsilon^{3/2}P(\epsilon)\nonumber\\
&-&\frac{\Omega}{2\sqrt{2}}\left(
\alpha^2\beta + \frac{3\beta}{2 \epsilon_F^{3/2}}\int
d\epsilon\,\sqrt\epsilon\,A(\epsilon) + \rm{c.c.}
\right).
\label{ENPERPART}
\eea
Using Eqs.~\eq{ALPHAEQ}--\eq{PEQ} it may be shown
straightforwardly that, provided one is willing to subtract certain formally equal
divergent integrals to obtain zero, the energy per particle
$\hbar e$ from Eq.~\eq{ENPERPART} is also a constant of the motion. The
divergent integrals are part of the issue of renormalization, which was
already discussed above and will be revisited again shortly.

We have made use of Eqs.~\eq{ALPHAEQ}--\eq{NORMEQ} in different 
variations, notations, and approximations many a time in the past in our
boson theories~\cite{JAV99,JAV02,MAC02,MAC02y}. Similarly, we have repeatedly
resorted to basically the same approach in the theory of the conversion between a
two-species Fermi gas and the corresponding diatomic
molecules~\cite{JAV04,JAV05}. Besides the obvious absence of the fermionic condensate, the difference is that where the boson
problem shows the factor
$1+2P(\epsilon)$ for Bose enhancement in Eq.~\eq{BETAEQ}, the corresponding fermion equation has
the factor
$1-2P(\epsilon)$ reflecting the exclusion principle.

\subsection{Pairing approximation for steady state}
Barring circumstances such as interference of the atomic BEC component of
the atom-molecule system with another reference BEC, a multiplicative
complex phase factor in the quantities $\alpha$, $\beta$ and $A(\epsilon)$ is
not observable. Besides, it is obvious from the equations of
motion~\eq{ALPHAEQ}--\eq{PEQ} that a certain combination of exponentially
evolving phases is self-sustained. Specifically, we search a
stationary solution in the form~\cite{JAV05}
\bea
\alpha(t) \equiv e^{-i\mu t}\alpha,\,\beta(t)\equiv e^{-2i\mu
t}\beta,\nonumber\\ A(\epsilon,t)\equiv A(\epsilon) e^{-2 i\mu t},
P(\epsilon,t)\equiv P(\epsilon)\,,
\label{ANSATZ}
\eea
where $\mu$ is a real frequency. It will turn out that $\hbar\mu$ is the
chemical potential for the atoms in this system (and half of the chemical
potential for the molecules), but such an interpretation is not
a given at this stage. In the rest of the paper we will again ignore the $\hbar$, and call $\mu$ the chemical potential.

Now, by a suitable choice of the zero of time we
may always make the coefficient
$\beta$ in Eqs.~\eq{ANSATZ} real and non-negative,
$\beta\ge0$; let us assume so from now on. To keep Eq.~\eq{PEQ} valid
with an Ansatz of the form~\eq{ANSATZ} at all times is then only possible if
($\beta=0$ or if)
$A(\epsilon)$ is real. Likewise, by Eq.~\eq{ALPHAEQ}, the amplitude
$\alpha$ must be real. With these restrictions, the time independent
coefficients must satisfy
\bea
\alpha\left(\mu+\frac{\beta\Omega}{\sqrt2} \right)&=&0\,,\label{ALPHAEQ0}\\
(2\mu-\delta)\beta&=&-{\Omega\over\sqrt{2}}\,\left[\alpha^2
+\frac{3}{2\epsilon_F^{3/2}}\int
d\epsilon\,\sqrt\epsilon\, A(\epsilon)\right]\,,\label{BETAEQ0}\nonumber\\
\\
(\mu-\epsilon) A(\epsilon) &=&
-\frac{\Omega}{2\sqrt{2}}[1+2P(\epsilon)]\beta\,,
\label{AEQ0}
\eea
and, of course, the norm condition~\eq{NORMEQ}.

The unknowns in the steady state are $\mu$, $\alpha$, $\beta$,
$A(\epsilon)$, and $P(\epsilon)$. Thinking about a numerical solution
with a discrete set of values for $\epsilon$, it is clear that there are
many more unknowns than equations. The problem is the original
Eq.~\eq{PEQ}, which will not lead to any useful
relation between $A(\epsilon)$ and $P(\epsilon)$ in the steady state.
Additional conditions are needed to constrain the solution.

The same dilemma came up in a system of two fermion species combining
into bosonic molecules. We resolved it~\cite{JAV05} by the assumption that the
fermions only come in pairs with opposite momenta and spins, as from
dissociation of molecules. This leads immediately to the relation between
pairing amplitudes and occupation numbers
\begin{equation}
|\langle c_{\bk\uparrow} c_{-\bk\downarrow}  \rangle|^2 =
 \langle c^\dagger_{\bk\uparrow}c_{\bk\uparrow} \rangle
-\langle c^\dagger_{\bk\uparrow}c_{\bk\uparrow}
\rangle^2\,,
\end{equation}
or, in the notation of the mean-field theory in Ref.~\cite{JAV05},
\begin{equation}
|C(\epsilon)|^2 -[P(\epsilon)-P^2(\epsilon)] = 0\,.
\label{FERMICONST}
\end{equation}
$C(\epsilon)$ is the fermion pairing amplitude analogous to
$A(\epsilon)$ of the present paper. With Eq.~\eq{FERMICONST}, the
number of equations was sufficient for a (presumably) unique
solution. Moreover, while we have not mentioned this before, the left-hand
side of Eq.~\eq{FERMICONST} is a constant of the motion in our BCS style
mean-field theory for fermions~\cite{JAV05}.

To address the corresponding boson case, let us take two momentum
states $\pm$, short for $\pm \bk$, with the occupation numbers
$|n_+n_-\rangle$. The most general completely paired state is of the form
\begin{equation}
|\psi\rangle = \sum_n c_n |n,n\rangle\,,
\label{PAIRSTATE}
\end{equation}
with $\sum_n |c_n|^2=1$. Given the propensity of boson to Poissonian statistics, we take
\begin{equation}
c_n = e^{-\half |\alpha|^2}\frac{\alpha^n}{\sqrt{n!}}\,,
\label{POISSPAIR}
\end{equation}
where $\alpha$ is a complex number. In the limit $|\alpha|\ll 1$ this
is a generic description for a situation when only the states
$|00\rangle$ and $|11\rangle$ are occupied, and the latter with a much
smaller probability. Similarly, in the limit of a real
$\alpha\equiv x \gg 1$ we have a generic description of the state in which
$|c_n|$ peak around $n\simeq x^2$, and $c_n$ vary slowly as a function of
$n$ around the maximum. In fact, we cover the case $|\alpha|\ll1$,
too, if we just use a real and positive $x$ in our argument, so that is
how we proceed.
Given the model, we have the expectation values
\bea
P&\equiv& \langle a^\dagger_+ a_+\rangle = x^2\,,\label{PNV}\\
A&\equiv& \langle a_+ a_-\rangle = e^{-x^2} x\sum_n
\frac{x^{2n}\sqrt{n+1}}{n!}\,\label{ANV}.
\eea

\begin{figure}
\begin{center}
\includegraphics[width=8.5cm]{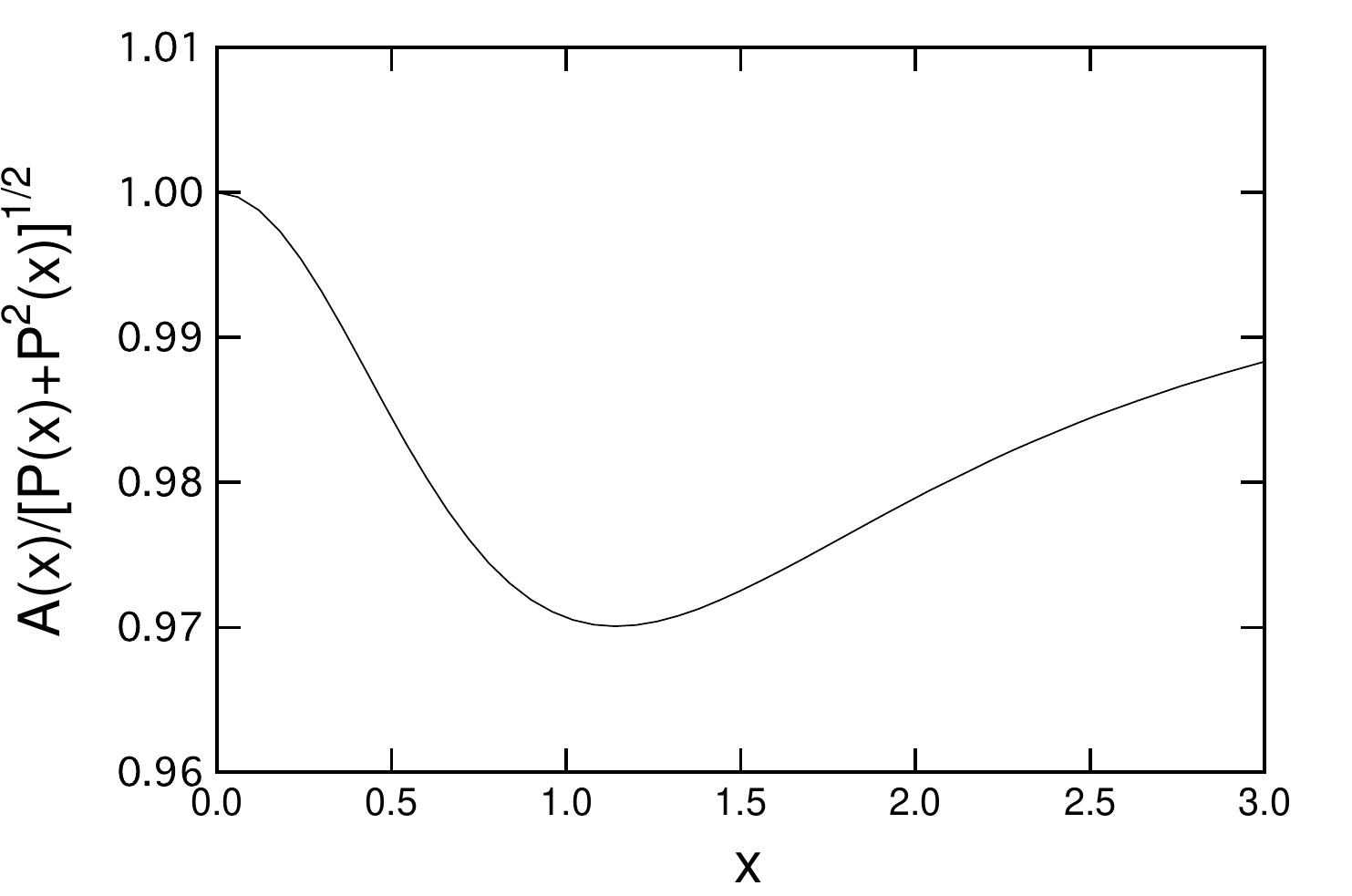}
\end{center}
\caption{The ratio of the actual pairing amplitude $A(x)$ and the
quantity $[P(x)+P^2(x)]^{1/2}$ derived from the occupation number of the
boson state $P(x)$ for the Poissonian paired
state, \protect\eq{PAIRSTATE} with~\protect\eq{POISSPAIR}, as a function of
the real parameter of the state $x=\alpha$. If the pairing
approximation\protect\eq{BOSECONST} were exact, this ratio would identically
equal unity.}
\label{PAIRING_ANSATZ}
\end{figure}

Mindful of the sign differences between bosons and fermions, on the basis
of Eq.~\eq{FERMICONST} we expect a relation for bosons of the form
\begin{equation}
|A(\epsilon)|^2-[P(\epsilon)+P^2(\epsilon)] = 0\,.
\label{BOSECONST}
\end{equation}
To find out if it works, we plot in Fig.~\ref{PAIRING_ANSATZ} the ratio
$A/\sqrt{P+P^2}$  from Eqs.~\eq{PNV} and~\eq{ANV} as a function of the
variable $x$. The maximum deviation of
this ratio from unity is about three per cent. We surmise that
\eq{BOSECONST} is a reasonable approximation between pairing amplitudes
and occupation numbers for boson states of the form~\eq{PAIRSTATE}.

The pairing approximation for bosons is further cemented by the observation
that, fully analogously to the fermion theory, in our mean-field theory for
bosons the left-hand side of~\eq{BOSECONST} is a constant of the motion.
This also gives an interesting piece of insight into the problem of finding
the steady state: There is a large (infinite) number of constants of the
motion, so that time evolution cannot lead to a unique steady state
without an explicit specification of the values of the constants. Physically,
on the other hand, it is generally up to the interactions with the
environment to force specific values, such as the zero on the right-hand
side of Eq.~\eq{BOSECONST}. In the corresponding fermion problem the zero
would follow even without the interactions with the environment if the system
started out as a condensate of  molecules, and the same
holds for the boson system. As advertised, our stationary solution
corresponds to the assumption that all noncondensate atoms are correlated as
if they came from dissociation of molecules.

Given that $A(\epsilon)$ is real and that it must be positive in the limit
$\epsilon\rightarrow\infty$ by virtue of Eq.~\eq{AEQ0},
Eqs.~\eq{AEQ0} and~\eq{BOSECONST} may be solved for the pairing amplitudes and
occupation numbers,
\bea
A(\epsilon) &=& \frac{\beta  \Omega }{2 \sqrt{2 (\epsilon -\mu )^2
-\beta^2\Omega ^2}}\label{AEPSILON}\,,\\
P(\epsilon) &=& \frac{1}{2} \left(\sqrt{\frac{\beta ^2 \Omega ^2}{2
(\epsilon -\mu )^2-\beta ^2 \Omega ^2}+1}-1\right)\label{PEPSILON}\,.
\eea
Hence, given the chemical potential $\mu$ and the amplitude of the
molecular condensate $\beta$, both the occupation numbers $P(\epsilon)$ and
pairing amplitudes $A(\epsilon)$ are uniquely determined. The inequality
\begin{equation}
\mu\le-\frac{\beta\Omega}{\sqrt{2}}
\label{MUINEQ}
\end{equation}
must hold, else complex occupation numbers would result. The equality in
Eq.~\eq{MUINEQ} presents no problem, since the ensuing singularities in the
occupation numbers and pairing amplitudes are sufficiently mild not to hamper
the analysis.
\subsection{Renormalization}
Equations~\eq{ALPHAEQ0}, \eq{BETAEQ0} and~\eq{NORMEQ} suffice to
determine the remaining unknowns $\alpha$, $\beta$ and $\mu$, although a few
issues remain. Next we discuss renormalization.

Consideration of the form of $A(\epsilon)$ in Eq.~\eq{AEPSILON} shows
right away that the integral $\int
d\epsilon\,\sqrt{\epsilon}\,A(\epsilon)$ in Eq.~\eq{BETAEQ0} diverges.
However, it turns out that the same renormalization that we devised for
the two-atom case also resolves this divergence. We replace $A(\epsilon)$ with
\begin{equation}
\bar{A}(\epsilon) = \frac{\beta  \Omega }{2 \sqrt{2 (\epsilon -\mu )^2
-\beta^2\Omega ^2}}-\frac{\beta  \Omega }{2 \sqrt{2\epsilon}}\,,
\label{RENSUB}
\end{equation}
which makes the integral convergent. But, to  keep Eq.~\eq{BETAEQ0} valid, we
need to add the the divergent integral $\int_0^\infty d\epsilon\, \epsilon^{-1/2}$
with an appropriate factor to the left-hand side as well. It turns out that
the net effect is precisely to replace the detuning $\delta$ on the left-hand
side with the renormalized detuning $\bar\delta$. Here we have played fast
and loose with mathematical rigor, but this could be remedied by 
introducing the upper limit $M$ to the integration just as in Eq.~\eq{UPLIM}
and then letting $M\rightarrow\infty$. 

Given the occupation numbers~\eq{PEPSILON}, the integral in the
normalization equation~\eq{NORMEQ} converges as written, but not so the
integral involving $P(\epsilon)$ in the expression for the energy per
particle~\eq{ENPERPART}. However, if one analogously to
Eq.~\eq{RENSUB} replaces $P(\epsilon)$ with
\begin{equation}
\bar{P}(\epsilon) = P(\epsilon) - \frac{(\beta \Omega)^2}{8 \varepsilon ^2}\,,
\end{equation}
in the integral involving $A(\epsilon)$ does the
subtraction~\eq{RENSUB}, {\em and\/} replaces the detuning with the
renormalized value $\bar\delta$, all divergences in Eq.~\eq{ENPERPART}
cancel. We regard this as an impressive demonstration of the consistency of
the mean-field theory.

\subsection{Statement of numerical problem}

We now have deal with the equations
\begin{widetext}
\bea
&&\alpha\left(\!\mu+\frac{\beta\Omega}{\sqrt2} \!\right)\!\!=\!\!0,\,
(2\mu-\bar\delta)\beta +
{\Omega\over\sqrt{2}}\left[\!\alpha^2
\!+\!\frac{3 (\beta  \Omega )^{3/2}}{4 2^{3/4}
\epsilon_F^{3/2}}
A_{1/2}\!\!\left(\!\!-\frac{\sqrt{2}\mu}{\beta\Omega}\!\right)\!\right]\!\!=\!\!0,\,
\alpha^2\!+\!\beta^2\!+\! 
\frac{3 (\beta  \Omega )^{3/2}}{4 2^{3/4} \epsilon _F^{3/2}}
P_{1/2}\!\!\left(\!\!-\frac{\sqrt{2}\mu}{\beta\Omega}\right)\!=
\!1;\label{TOSOLVE}\\
&&e =
 \frac{\beta ^2\bar \delta }{2} - \frac{\beta  \Omega }{\sqrt{2}}
\left\{
\alpha^2 + 
\frac{3 (\beta  \Omega )^{3/2}}{4 2^{3/4} \epsilon_F^{3/2}}
\left[
A_{1/2}\left(-\frac{\sqrt{2} \mu }{\beta  \Omega
}\right)-P_{3/2}\left(-\frac{\sqrt{2} \mu }{\beta  \Omega }\right)
\right]
\right\};\label{ENERGY}\\
&&A_{1/2}(m) = \int_0^\infty dx\,\frac{-m^2-2 x
m+1}{\sqrt{x}\sqrt{[(m+x)^2-1]}
\left(x+\sqrt{(m+x)^2-1}\right)};\label{AHALF}\\
&&P_{1/2}(m) = \int_0^\infty dx\,\frac{\sqrt{x}}{\sqrt{(m+x)^2-1}
\left(m+x+\sqrt{(m+x)^2-1} \right)};\label{PHALF}\\
&&P_{3/2}(m) =  \int_0^\infty dx\, \frac{-8 m x^3-4 m^2 x^2+3 x^2-2 m
x-m^2+1}{2 \sqrt{x} \sqrt{(m+x)^2-1} 
\left[2(x+m)x^2+(2x^2+1)
\sqrt{(m+x)^2-1}\right]}\,.\label{PTHREEHALVES}
\eea
\end{widetext}
Equations~\eq{TOSOLVE} are the ones to solve for the unknown quantities
$\alpha$, $\beta$ and $\mu$, and Eq.~\eq{ENERGY} gives the resulting
mean-field energy per particle. The integral $A_{1/2}(m)$ is a representation
of the integral $\int d\epsilon\,\epsilon^{1/2} \bar{A}(\epsilon)$ as a function of
the chemical potential $\mu$, and similarly for $P_{1/2}$, $P_{3/2}$. These
integrals are properly renormalized and dimensionless, and we have gone so
far as to write them in forms that do not involve near-canceling subtractions
of large numbers. They are suitable for use in numerical
computations as written.
\section{Solving the theory}
We have solved Eqs.~\eq{TOSOLVE}-\eq{PTHREEHALVES} using {\em
Mathematica\/}~\cite{MATH} in a combination of analytical and numerical
calculations. We
double-checked many of the results by independent programming on
Maple~\cite{MAPLE}. Unlike in the fermion case~\cite{JAV05} where the production of
accurate numerical results for arbitrary parameter values was a major project
in numerical analysis, with bosons we never had to resort to a general-purpose programming
language such as C++. 

In detailed studies we first investigated numerically how the results behave, and
use this knowledge to formulate Ansatz solutions to find analytical results.
The discussion preceding Eqs.~\eq{LINEASYMPTOTICS} below serves as an
example. An enormous amount of detail could be extracted in this way, but
our aim is to demonstrate a few major qualitative features only.

\subsubsection{Atomic condensate present}
One way of satisfying Eq.~\eq{ALPHAEQ0}, or the same equation as the
first member of Eqs.~\eq{TOSOLVE}, is to require that the expression
inside the brackets vanishes. This leads to
\begin{equation}
\mu = -\frac{\beta\Omega}{\sqrt2}\,,
\label{DEFMU}
\end{equation}
the integrals become $A_{1/2}(1) = -2\sqrt{2}$, $P_{1/2}(1) = 2\sqrt2/2$,
and $P_{3/2}(1)= - 8 \sqrt2/5$, and the equations to solve for $\alpha$ and
$\beta$ are
\bea
(\bar\delta -2 \mu )\beta =\frac{\alpha ^2 \Omega }{\sqrt{2}}-\frac{3 \beta
^{3/2}
\Omega^{5/2}}{2^{7/4}\, \epsilon _F^{3/2}}\,,\label{DEFBETA}\\
\alpha ^2+\beta ^2+\frac{(\beta  \Omega )^{3/2}}{2^{5/4} \epsilon
_F^{3/2}}=1\,\label{DEFNORM}.
\eea
In principle these have an explicit closed-form solution, but in practice we
have found it useless and proceed numerically. Given the solution, the energy
per particle may be found from
\begin{equation}
e=\frac{3 (-\mu )^{5/2}}{5 \sqrt{2}\,\epsilon_F^{3/2}}+\alpha ^2 \mu\,.
\label{EANE0}
\end{equation}

Now, by eliminating $\alpha$ from Eqs.~\eq{DEFBETA} and~\eq{DEFNORM} we
have a necessary condition for the solution,
\begin{equation}
\beta ^2+\frac{\sqrt{2} \left(\sqrt{2} \beta  \Omega+\bar{\delta }\right)
\beta }{\Omega }+\frac{2^{3/4} (\beta  \Omega )^{3/2}}{\epsilon _F^{3/2}}=1\,.
\end{equation}
It is easy to see numerically that
for any real $\bar\delta$ and $\Omega>0$ this has at most one solution with
$0\le\beta\le1$. Moreover, numerically one may
demonstrate that such a solution
$\beta(\Omega,\bar\delta)$ is a decreasing function of $\bar\delta$.
Therefore, a lower limit on the existence of a solution as a function of
$\bar\delta$ is a possibility: Once
$\beta$ has reached a certain value $<1$, Eq.~\eq{DEFNORM} dictates that
$\alpha^2=0$. Suppose one decreases $\bar\delta$ further, then $\beta$ would
increase further and Eq.~\eq{DEFNORM} would require
$\alpha^2<0$, which is not allowed for a real $\alpha$.

\begin{figure}
\begin{center}
\includegraphics[width=8.5cm]{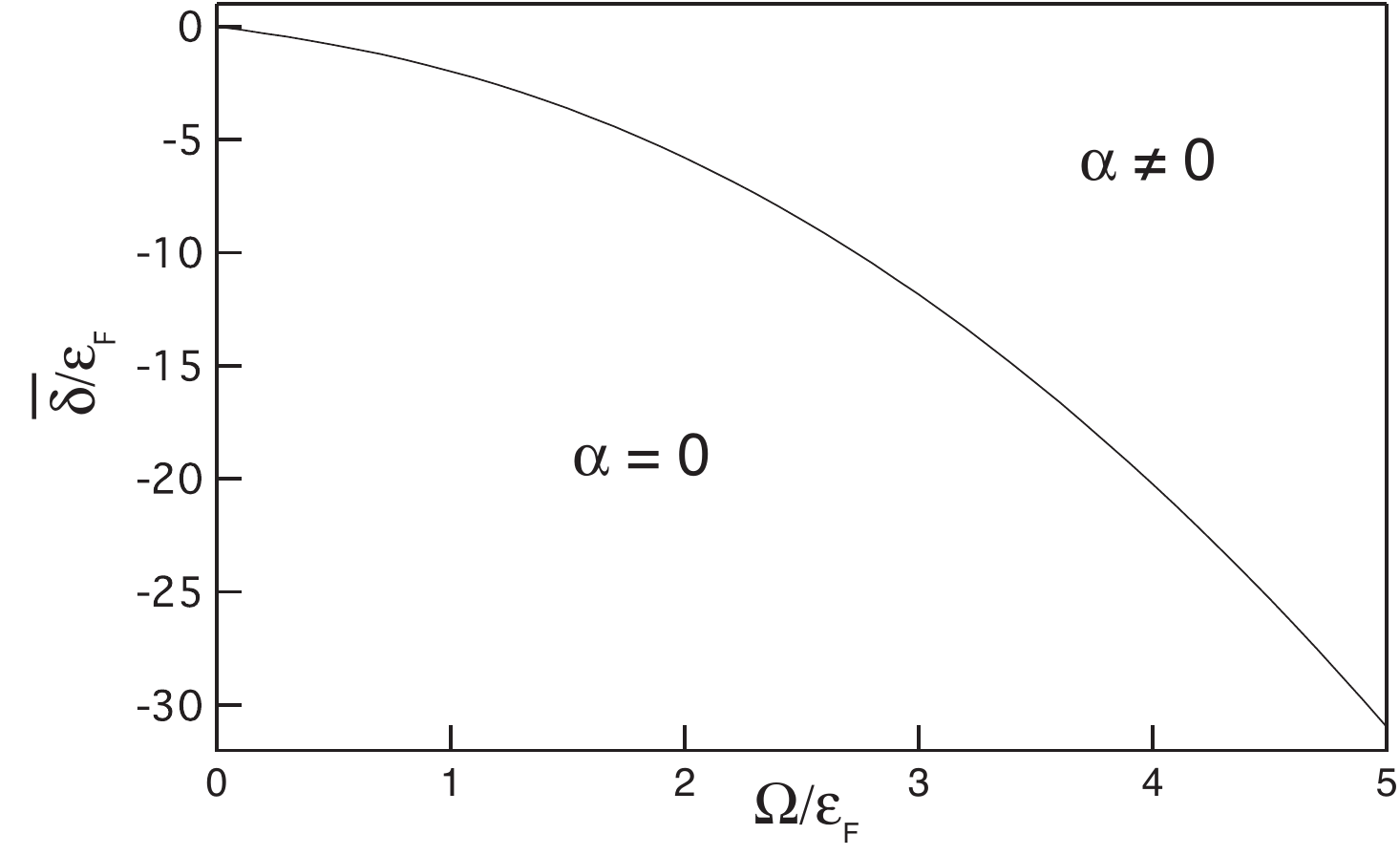}
\end{center}
\caption{The line in the $(\Omega,\bar\delta)$
plane separating the phases of the system with a BEC of atoms
present ($\alpha\ne0$) and absent ($\alpha=0$).}
\label{PT_LINE}
\end{figure}

The limiting case
is found setting $\alpha=0$ in Eqs.~\eq{DEFMU}--\eq{DEFNORM}, which gives a
relation between $\bar\delta$ and $\Omega$. We plot in Fig.~\ref{PT_LINE} the
detuning $\bar\delta$ obtained numerically in this way as a function of the
Rabi frequency $\Omega$, effectively using the Fermi frequency as a scale of
frequencies. Given the equations to solve, we may also attempt power series
solutions. For instance, we may insert the series
\bea
-\bar\delta &=& K_1 \Omega^1 + K_{3/2} \Omega^{3/2}+K_{2} \Omega^{2} +
\ldots\,,\\
-\mu &=& M_1 \Omega^1 + M_{3/2} \Omega^{3/2}+M_{2} \Omega^{2} +
\ldots\,,
\eea
into Eqs.~\eq{DEFMU}--\eq{DEFNORM}, and attempt to satisfy the equations power
by power in $\Omega$. This Ansatz works and gives solvable equations for
the coefficients $K$ and $L$. Following such principles, we find the
expressions of the curve in Fig.~\ref{PT_LINE} for the limits of
both small and large $\Omega$,
\begin{equation}
\bar\delta(\Omega) = \left\{
\begin{array}{ll}
-\sqrt{2}\,\Omega+{\cal O}(\Omega^{5/2}),&\Omega\ll\epsilon_F\,,\\
-\frac{\displaystyle3}{\displaystyle2^{4/3}}\,\frac{\displaystyle\Omega^2}{\displaystyle\epsilon_F}+{\cal
O}(\Omega^{0}),&\Omega\gg\epsilon_F\,.
\end{array}
\right.
\label{LINEASYMPTOTICS}
\end{equation}

\subsubsection{Atomic condensate absent}
The other way to satisfy Eq.~\eq{ALPHAEQ0} is to set
\begin{equation}
\alpha=0\,,
\end{equation}
which corresponds to the up-front statement that there is no atomic
condensate. Equations~\eq{TOSOLVE} then turn into
\bea
\mu &\equiv& -\frac{\beta\Omega}{\sqrt2}\,m\,,\label{A0MU}\\
\left({\bar\delta-2 \mu}\right)\beta  &=&\frac{3 \Omega  (\beta  \Omega
)^{3/2}
 A_{\frac{1}{2}}(m)}{2^{13/4} \epsilon _F^{3/2}}\,,\label{A0BTA}\\
1&=&\beta ^2+\frac{3 (\beta  \Omega )^{3/2} P_{\frac{1}{2}}(m)}{2^{11/4}
\epsilon _F^{3/2}}\,\label{A0NORM}\,,
\eea
which are to be solved for $\beta$
and $\mu$; we have expressed the chemical potential $\mu$ using the
dimensionless variable $m\in[1,\infty)$. The
integrals~\eq{AHALF}--\eq{PTHREEHALVES} may be written in terms of elliptic
integrals, but this fact appears to be useless and the practical
solutions again proceed numerically. Once the solution is found, the energy
per particle is given by Eq.~\eq{ENERGY} as written.

Now, consider
Eqs.~\eq{A0MU}--\eq{A0NORM} for a fixed value of the parameter $\Omega$,
regarding $\beta$ and $\bar\delta$ as functions of the variable $m$.
By plotting the respective functions, it may be seen that, for
$m>1$, $P_{1/2}(m)>0$,
$P'_{1/2}(m)<0$, $A_{1/2}(m)<0$, and $A'_{1/2}(m)<0$, with~$'$ denoting the
derivative. Equation~\eq{A0NORM} then implies that
$\beta'(m) >0$, and Eq.~\eq{A0BTA} consequently gives  $\bar\delta'(m)<0$.
In other words, when the parameter $\bar\delta$ is increased while keeping
$\Omega$ fixed,  $m$ and $\beta$ resulting from
Eqs.~\eq{A0MU}--\eq{A0NORM} decrease. But, by the time the detuning
$\bar\delta$ has reached the line in Fig.~\ref{PT_LINE}  the parameter
$m$ has attained the minimum permissible value $m=1$, and there cannot
be a solution for any larger $\bar\delta$.

\subsubsection{Role of atomic condensate}
The two classes of solutions we have found, $\alpha>0$ and
$\alpha\equiv0$ corresponding to the presence and absence of an atomic
condensate join continuously. On the line drawn in
Fig.~\ref{PT_LINE}
$\alpha=0$ and
$\mu=-\beta\Omega/\sqrt{2}$ both hold true, so that all of the
Eqs.~\eq{DEFMU}--\eq{DEFNORM} as well as Eqs.~\eq{A0MU}--\eq{A0NORM} are
satisfied simultaneously.

Summarizing, we have the following observations about the structure
of the theory. There are two different kind of solutions for $\mu$, $\alpha$
and
$\beta$ when the parameters $\Omega$ and $\bar\delta$ are varied,
characterized by the conditions
$\mu=-\beta\Omega/\sqrt{2}$ and $\alpha=0$. Only the former (latter) exists
in the region of parameters  $\Omega$ and $\bar\delta$ labeled $\alpha\ne0$
($\alpha=0$) in Fig.~\ref{PT_LINE}. On the borderline between the regions both
solutions exist and agree, so that they go continuously from one to the other
as the variable $\bar\delta$ and/or $\Omega$ crosses the line. The solution is unique for all
$\bar\delta$ and $\Omega$ under the assumptions $\Omega>0$, $\alpha\ge0$
and $\beta>0$, which are a matter of convenience and can always be made.
Finally, our numerical computations demonstrate that the unique solution
always exists.

\section{Features of the theory}
The present boson theory has three parameters with the dimension of frequency,
$\Omega$, $\bar\delta$, and $\epsilon_F$. In our discussions we regard
Fermi energy, $\epsilon_F$,  as the scale of frequencies,
although we always write it down explicitly. In present-day dilute quantum
degenerate gases the representative value is $\epsilon_F\sim
2\pi\times 10\,\text{kHz}$. In the classic Feshbach resonance experiments
with bosons orders of magnitude of  the coupling Rabi frequencies $\Omega\sim10\epsilon_F$ are
typical~\cite{JAV02}, but a narrower Feshbach resonance could change this
comparison significantly. In photoassociation the ratio $\Omega/\epsilon_F$
may be varied by varying the intensity (intensities) of the
laser(s). In a Feshbach resonance the detuning $\bar\delta$ depends on the
product of magnetic field and the difference of the magnetic dipole moments
of the bound molecular state and of the free-atom state. Roughly,
$\bar\delta\sim\epsilon_F$ corresponds to 10~mG change in the magnetic field. In
photoassociation the parameter $\bar\delta$ is directly a matter of tuning of
the laser(s).

For comparison it is useful to recall a few basic facts about the usual one-channel theory for a BEC~\cite{STR99,KOH06}. In such modeling there are no explicit molecules at all, but it is assumed that the atoms interact among themselves as characterized by a scattering length $a$ that may be tuned by varying the magnetic field around the Feshbach resonance, i.e., by varying the detuning. At positive detunings the scattering length is negative, which indicates an attractive interaction between the atoms and collapse of the (untrapped) condensate. At negative detunings, at least close to the resonance, the scattering length is positive and large. According to the theory of molecular structure, this means that there must be a weakly bound molecular state, which, however, is not included explicitly in the one-channel theory. By elementary thermodynamics, at low temperatures these molecules should make the thermal-equilibrium state. Positive (negative) detunings thus indicate the atom (molecule) side of the resonance.

\subsection{Scaling with detuning and coupling strength}

\begin{figure}
\begin{center}
\includegraphics[width=8.5cm]{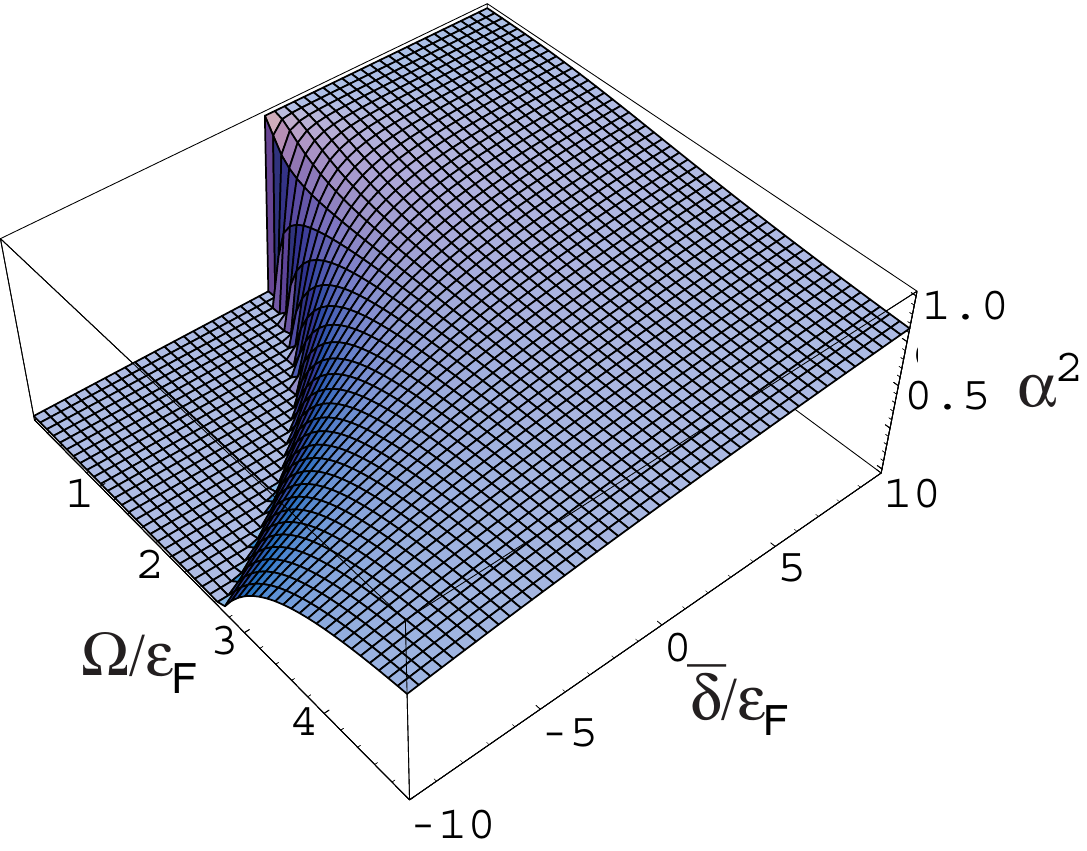}
\end{center}
\caption{(Color online). Fraction of atoms in the atomic BEC, $\alpha^2$, as a function of
the coupling strength, $\Omega$, and detuning, $\bar\delta$.}
\label{3D_ASQ}
\end{figure}

In Fig.~\ref{3D_ASQ} we plot the fraction of atoms in the atomic BEC,
$\alpha^2$, as a function of the coupling strength $\Omega$ and detuning
$\bar\delta$. At $\Omega\ll \epsilon_F$, the system switches abruptly
(compared to
$\epsilon_F$) from having all atoms to no atoms in the condensate when the
detuning crosses the two-body resonance position
$\bar\delta=0$. As the coupling $\Omega$ increases, the transition from
$\alpha^2\simeq1$ to $\alpha^2\ll1$ rounds out. However, $\alpha^2=0$ is
always reached on the line shown in Fig.~\ref{PT_LINE}. Given
that on this line the solution of the system is  not analytic, we predict a
phase transition.

\begin{figure}
\begin{center}
\includegraphics[width=8.5cm]{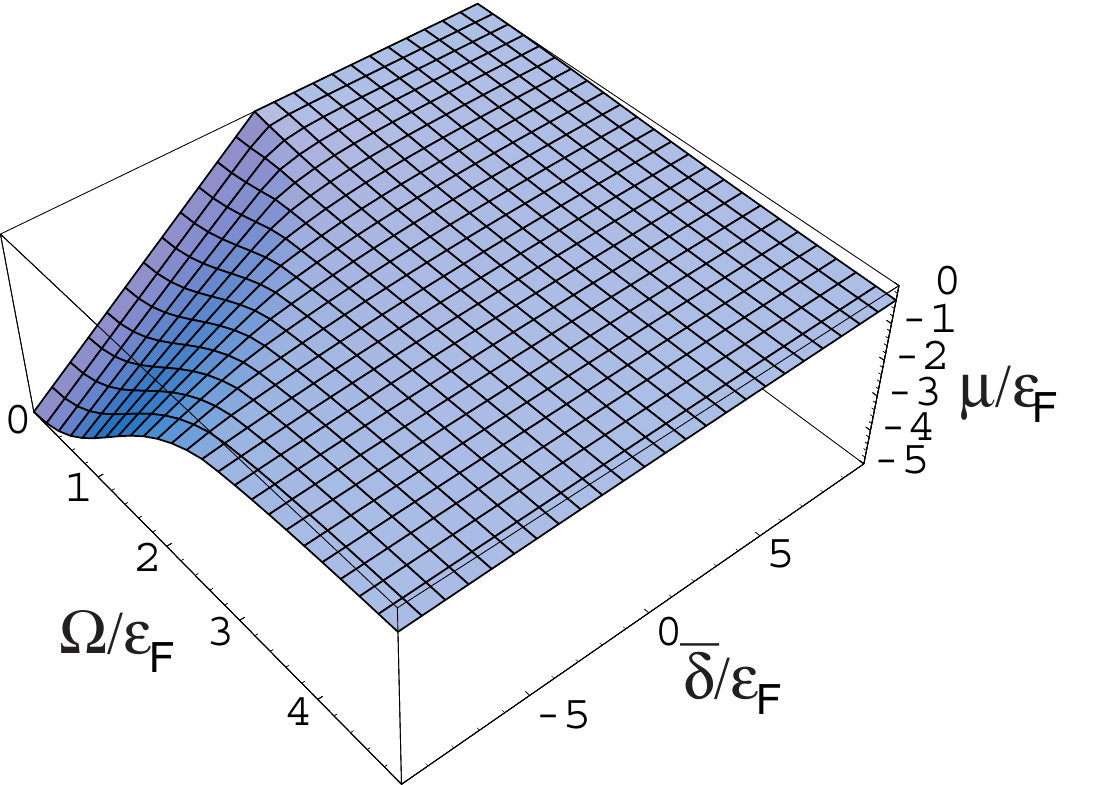}
\end{center}
\caption{(Color online). Chemical potential $\mu$ as a function of
the coupling strength $\Omega$ and detuning $\bar\delta$.}
\label{3D_MU}
\end{figure}

The analogous plot for the chemical potential $\mu$ as a function of coupling
strength and detuning is given in Fig.~\ref{3D_MU}. The function
$\mu(\Omega,\bar\delta)$ is also nonanalytic on the line shown in
Fig.~\ref{PT_LINE}, but the kink is not visible
in Fig.~\ref{3D_MU}. The rounding-out with increasing coupling strength is again obvious.

\begin{figure}
\begin{center}
\includegraphics[width=8.5cm]{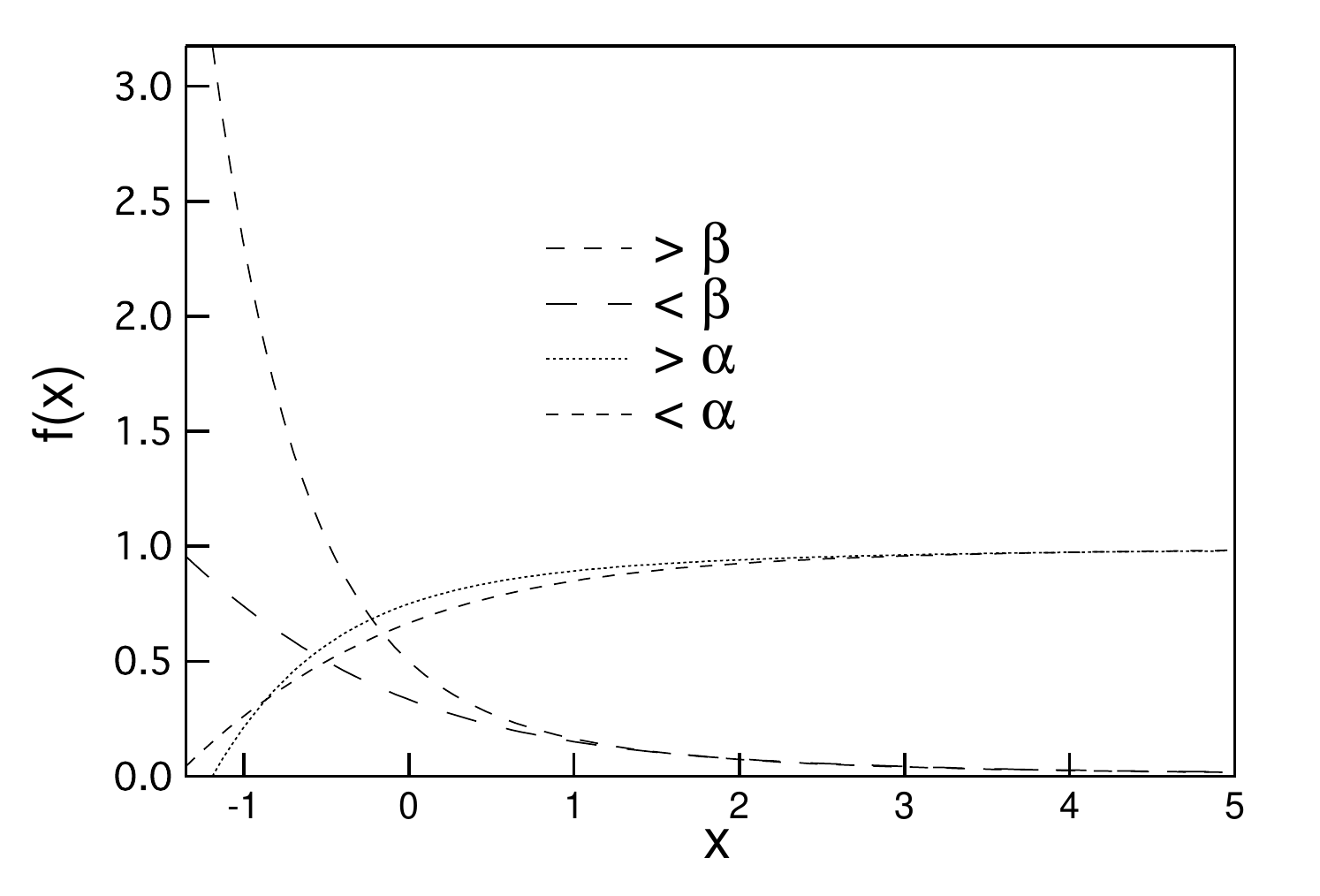}
\end{center}
\caption{Scaling functions $f^>_\beta(x)$, etc., in
Eqs.~\protect\eq{ASCALING} and~\protect\eq{BSCALING}. The legend identifies
the combination of $<$ or $>$ and $\alpha$ or $\beta$ for each curve.}
\label{SQ_FUNCTIONS}
\end{figure}

We quantify the rounding-out starting with the numerical observation that in the regime
$\alpha\ne 0$, the following limits hold true:
\begin{eqnarray}
\alpha^2(\Omega,\bar\delta)&=&\left\{
\begin{array}{ll}
f^<_\alpha\left[\frac{\displaystyle\bar\delta}{\displaystyle\Omega}\right]\,,&
\Omega\ll\epsilon_F\,;\\
f^>_\alpha\left[\left(\frac{\displaystyle\bar\delta}{\displaystyle\Omega}
\right)^2\right]\,,&\Omega\gg\epsilon_F\,;
\end{array}
\right.\label{ASCALING}\\
\beta^2(\Omega,\bar\delta)&=&\left\{
\begin{array}{ll}
f^<_\beta\left[\frac{\displaystyle\bar\delta}{\displaystyle\Omega}\right]\,,
&\Omega\ll\epsilon_F\,;\\
\left(\frac{\displaystyle\epsilon_F}{\displaystyle\Omega}\right)^2 \,
f^>_\beta\left[\left(\frac{\displaystyle\bar\delta}{\displaystyle\Omega}
\right)^2\right]\,,&\Omega\gg\epsilon_F\,.
\end{array}
\right.\label{BSCALING}
\end{eqnarray}
Knowing these scalings it is easy to work out analytical expressions
for the functions $f^<_\alpha$, etc., from Eqs.~\eq{A0MU}--\eq{A0NORM};
simply put in the scalings, and see what becomes of the equations in the
corresponding limit of $\Omega$. For instance, we have
$f^>_\beta(x)=[y(x)]^2$, where
$y$ is a solution to the polynomial equation
\begin{equation}
8 y^3-4 \sqrt{2}\, x^2 y^2+8 x y-2 \sqrt{2}=0\,
\end{equation}
for the fixed value of $x$. We have plotted the scaling functions in
Fig.~\ref{SQ_FUNCTIONS}. They tend to the proper limits in both ends of the
range of the variable $x$, and, as appropriate for useful dimensionless scaling
functions, their magnitudes and scales of variation with the argument $x$
are both on the order of unity.

We next turn to the behavior of the chemical potential and the energy per
particle, particularly in the region $\alpha=0$ in Fig.~\ref{PT_LINE} with
both
$\bar\delta<0$ and $\mu<0$. To this end we first note the asymptotic expansions
of the functions~\eq{AHALF}--\eq{PTHREEHALVES} for $m\rightarrow\infty$:
\bea
A_{1/2}(m)&\sim& -\pi \sqrt{m}+\frac{\pi}{16}\, m^{-3/2}+{\cal
O}(m^{-7/2})\label{AHASY}\,,\\
P_{1/2}(m)&\sim& \frac{\pi}{4\sqrt m} + \frac{3\pi}{128}\,m^{-5/2} + {\cal
O}(m^{-9/2})\label{PHASY}\,,\\
P_{3/2}(m)&\sim&-\frac{3\pi}{4}\,\sqrt{m} +\frac{3\pi}{128}\,m^{-3/2} + {\cal
O}(m^{-7/2})\,.\label{PTHASY}\hspace{20pt}
\eea
The coefficients were inferred from numerical results as given by {\it
Mathematica}'s\ \ {\tt NIntegrate} function. They are probably exact; the
coefficients of the leading terms appear to be precise to at least $10^{-8}$,
and the coefficients in the next-to-leading terms to about $10^{-3}$.

Now, inserting power series expansions of the form
\bea
-\mu &=& \frac{|\bar\delta|}{2} +
M_{1/2}|\bar\delta|^{1/2}+M_{0}|\bar\delta|^{0}+\ldots\,,\label{MUASY}\\
\beta&=&1+B_{-1/2}|\bar\delta|^{-1/2}+B_{-1}|\bar\delta|^{-1}+
\ldots\label{BASY}\
\eea
into Eqs.~\eq{A0MU}--\eq{A0NORM} and expanding the result to a power series
in $1/|\bar\delta|$ using Eqs.~\eq{AHASY}--\eq{PTHASY}, we discover that we may determine the coefficients $M$ and
$B$ in Eqs.~\eq{MUASY} and~\eq{BASY} down to and including
$M_{-3}$ and
$B_{-4}$ in terms of the explicit coefficients in
Eqs.~\eq{AHASY}--\eq{PTHASY}. These, in turn, will give an expression for the
energy per particle~\eq{ENERGY} in the form
\begin{equation}
e=-\frac{\bar\delta}{2}+E_{1/2}|\bar\delta|^{1/2}+E_{0}|\bar\delta|^{0}+
\ldots\,,\label{EASY}
\end{equation}
also down to and including the order $|\bar\delta|^{-3}$.

We will not write down the expansions~\eq{MUASY}, \eq{BASY} and~\eq{EASY}
in detail, but a few notes are relevant. The expansions of chemical potential
and energy go with powers of the dimensionless number
$z=\Omega^4/\epsilon_F^3|\bar\delta|$ down to and including the order
$|\bar\delta|^{-1}$, and the expansion of condensate amplitude down to and
including the order $|\bar\delta|^{-2}$, although in the expansions of $\mu$
and $e$ the coefficients of $|\bar\delta|^{-1}$ happen to be zero,
$M_{-1}=E_{-1}=0$. But since we have
$\Omega\propto\rho^{1/2}=(N/V)^{1/2}$ and
$\epsilon_F\propto\rho^{2/3}=(N/V)^{2/3}$, the parameter $z$ does not depend on
density, or particle number, or volume, at all. It is a single-molecule
quantity. In fact, with the identification~\eq{CONTVSDISCR},
down to the order $|\bar\delta|^{-1}$ the expansions of $\mu$ and $e$ coincide
with the expansion of $\omega_b/2$, where $\omega_b$ is the bound-state
energy of a single molecule as given in Eq.~\eq{BSENE} in the Appendix.

Many-body phenomena start at the order
$|\bar\delta|^{-3/2}$, from which onwards other combinations of $\Omega$ and
$\epsilon_F$ than $z$ also emerge. For instance, asymptotically,
$|\bar\delta|\rightarrow\infty$, the dependence of chemical potential and
energy on detuning and density goes as
$\propto \rho/|\bar\delta|^{3/2}$. The analogous mean-field theory for
fermions~\cite{JAV05} behaves in this respect in the same way.

The conventional scattering length for atoms behaves
as $a\propto-\bar\delta^{-1}$ near the Feshbach resonance, and for fermions the scattering length of the molecules near the resonance should be $0.6$ times the
scattering length for the atoms~\cite{PET04a}. This means that the density
dependence of energy should emerge in the order $\bar\delta^{-1}$. The present
mean-field theory does not conform with the standard expectations.
Whether this is a contradiction or not, we cannot say. First, by
their very nature the asymptotic expansions are valid far away from
the resonance, not close where one expects the $0.6\,a$ for molecules made of fermions. Second,
as a practical matter, the frequency $\Omega^4/\epsilon_F^3$ is typically
quite large for a Feshbach resonance; we estimate $\sim 2\pi\times$10 GHz for
the usual Feshbach resonance in
${}^{85}$Rb~\cite{JAV06}, and indeed THz scale values for the 834~G Feshbach
resonance in fermionic ${}^6$Li~\cite{JAV05}. Our asymptotic expansions only
become useful at extremely large detunings, when in practice some other
physics assumption of our theory such as the contact interaction or the
neglect of the background scattering length has already become invalid. We do
not expect the expansions~\eq{MUASY}--\eq{EASY} to be of much practical
value, which is one reason why we have not listed the coefficients. We will
return to their theoretically interesting properties below, though.

We next turn to the limit of strong coupling,
$\Omega\rightarrow\infty$. Let us first fix $\bar\delta=0$, so that the
argument is within the regime $\alpha\ne0$. We set $\bar\delta=0$ in
Eqs.~\eq{DEFMU}--\eq{DEFNORM}, attempt a solution of the form
\bea
\alpha &=& A_0 + A_{-2} \Omega^{-2}+A_{-4}\Omega^{-4} + \ldots\,,\\
-\mu &=& M_0 + M_{-2} \Omega^{-2}+M_{-4}\Omega^{-4}+\ldots\,,
\eea
and find that it works; asymptotically, with $\Omega\rightarrow\infty$, we
have $\mu = -\half\epsilon_F$ and $\alpha^2=3/4$. From Eq.~\eq{EANE0}, the
corresponding expression for energy per particle is
$e=-\hbox{$1\over3$}\epsilon_F$. As expected on the basis of unitarity~\cite{BRU04,DIE04}, in the limit of very strong interactions the interaction strength vanishes from the result and the only
energy scale that remains, $\epsilon_F$, is set by the density of the gas.

Given that the line separating the $\alpha=0$ and $\alpha\ne 0$ regions in
Fig.~\ref{PT_LINE} may, perhaps, be thought of as the position of
the Feshbach resonance shifted by many-body effects, it is also instructive
to find the asymptotic limits of chemical potential and energy as one moves
along this line. To find them, we set $\alpha=0$ in
Eqs.~\eq{DEFMU}--\eq{DEFNORM} and put in the Ansatz
\bea
-\bar\delta&=& D_2\Omega^2 + D_{0}\Omega^{0} + D_{-2}\Omega^{-2} + \ldots\,,\\
-\mu &=& M_0 + M_{-2} \Omega^{-2}+M_{-4}\Omega^{-4}+\ldots\,.
\eea 
Aside from reproducing the second of Eqs.~\eq{LINEASYMPTOTICS}, this analysis shows that in the limit of large interaction strength we have
$\mu=-\sqrt[3]{2}\,\epsilon_F$. This, in turn, gives
$e=-3\sqrt[3]{2}/5\,\epsilon_F$.

\subsection{Thermodynamics}

The atom-molecule BEC has some interesting thermodynamical properties,
in particular a propensity for negative pressure. We now discuss these aspects.

It seems natural to identify the mean-field value of the energy $E = \hbar N
e$ as the thermodynamic internal energy $U$ of the system. Moreover,
the system is at zero temperature, which indicates an entropy identically
equal to zero. In the sense of thermodynamics, the chemical potential should
then equal
\begin{equation}
\mu_T = \left(\frac{\partial U}{\partial N}\right)_V\,.
\label{TDMUDEF}
\end{equation}

There is a subtlety to this definition, in that $N$ is the invariant atom number that combines both atoms and molecules, yet we have used it to find the chemical potential for atoms only. On the other hand, the condition for chemical equilibrium for atoms and molecules dictates that the thermodynamic chemical potentials for atoms and molecules satisfy $2\mu_{T,a}=\mu_{T,m}$. From this observation it is easy to deduce that the procedure~\eq{TDMUDEF} gives the correct chemical potential for the atoms. There are
related issues elsewhere in our thermodynamics discussions that could be resolved similarly, but henceforth we will not bring them up.

We have talked of the quantity $\mu$ as the chemical potential, which is
consistent with our thermodynamics identification $E\leftrightarrow U$ if 
\begin{equation}
\mu = \left(\frac{\partial [N e]}{\partial N}\right)_V\,.
\label{MUIDENTITY}
\end{equation}

As already noted repeatedly, both atom number and volume are ingredients
in our theory because the parameters $\Omega$ and $\epsilon_F$ depend on the
density; for instance,
\begin{equation}
\left(\frac{\partial \Omega}{\partial N}\right)_V = \frac{\Omega}{2N}, \quad
\left(\frac{\partial \epsilon_F}{\partial N}\right)_V =
\frac{2\epsilon_F}{3N}\,.
\label{DERIDS}
\end{equation}
However, verifying Eq.~\eq{MUIDENTITY} by simply taking the analytical $N$
derivative of Eq.~\eq{ENERGY} appears cumbersome,
since we should then take into account the implicit dependence on $N$ of the
quantities $\alpha$ and $\beta$, and even of $\mu$ itself. Instead, we have
carried out the verification numerically, by calculating the derivative of $e$
numerically using~Eqs.~\eq{DERIDS}, and found that Eq.~\eq{MUIDENTITY} is
indeed  satisfied.

We have also found that the analytical asymptotic
expansions~\eq{EASY} and~\eq{MUASY} satisfy Eq.~\eq{MUIDENTITY}
to all orders, down to and including $|\bar\delta|^{-3}$, for which the
expansion coefficients are explicitly known. This is a rather impressive
confirmation of our originally numerical identification of the expansion
coefficients in Eqs.~\eq{AHASY}--\eq{PTHASY}.

Finally, there is the question of pressure. The thermodynamical expression for
pressure in the present case with identically zero entropy is
\begin{equation}
p = - \left(\frac{\partial U}{\partial V}\right)_N = -\hbar N 
\left(\frac{\partial e}{\partial V}\right)_N\,.
\label{PDEF}
\end{equation}
But, given the identifications we have already made, the Gibbs-Duhem
relation of thermodynamics reads
\begin{equation}
\hbar  Ne = -pV + \hbar N\mu\,,
\end{equation}
or
\begin{equation}
\frac{p}{\hbar \rho} = \mu-e\,.
\label{PGD}
\end{equation}
We have calculated the pressure both ways numerically, too, and found that
Eqs.~\eq{PDEF} and~\eq{PGD} are consistent.

We have already noted that in the case
$\bar\delta=0$ and $\Omega\rightarrow\infty$, the limits $\mu
\rightarrow-\half\epsilon_F$ and $e
\rightarrow-\hbox{$1\over3 $}\epsilon_F$ hold true, which gives
$p/(\hbar\rho)\rightarrow -\hbox{$1\over6 $}\epsilon_F$. The pressure becomes
negative. In a way, this is as expected. Assuming that the transition line
between the $\alpha=0$ and $\alpha\ne0$ lines denotes the true Feshbach
resonance as shifted by the interactions in the system, this limiting case is
on the atom side the Feshbach resonance where, according to the standard one-channel
model, the scattering length is negative and the BEC of the atoms is liable to
collapse. This is also what one would expect for negative pressure.

\begin{figure}
\begin{center}
\includegraphics[width=8.5cm]{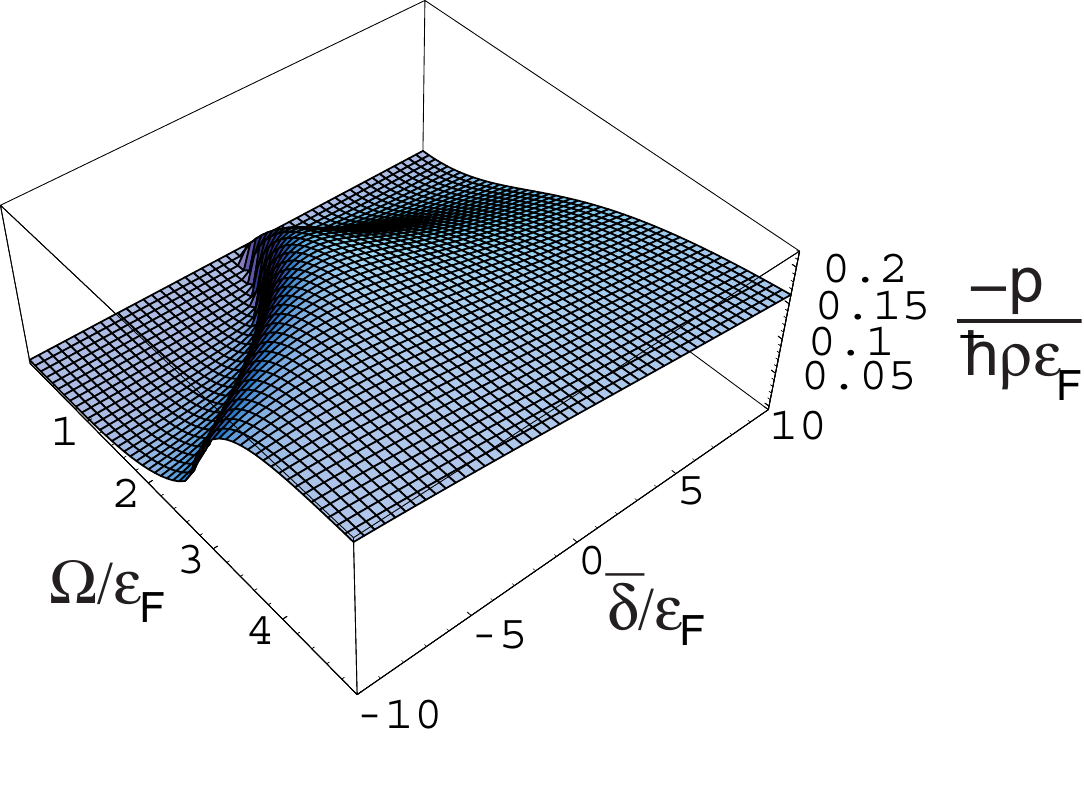}
\end{center}
\caption{(Color online). Pressure $p$ of the atom-molecule BEC plotted as a function of the
interaction strength $\Omega$ and detuning $\bar\delta$.}
\label{PRESSURE}
\end{figure}

\begin{figure}
\begin{center}
\includegraphics[width=8.5cm]{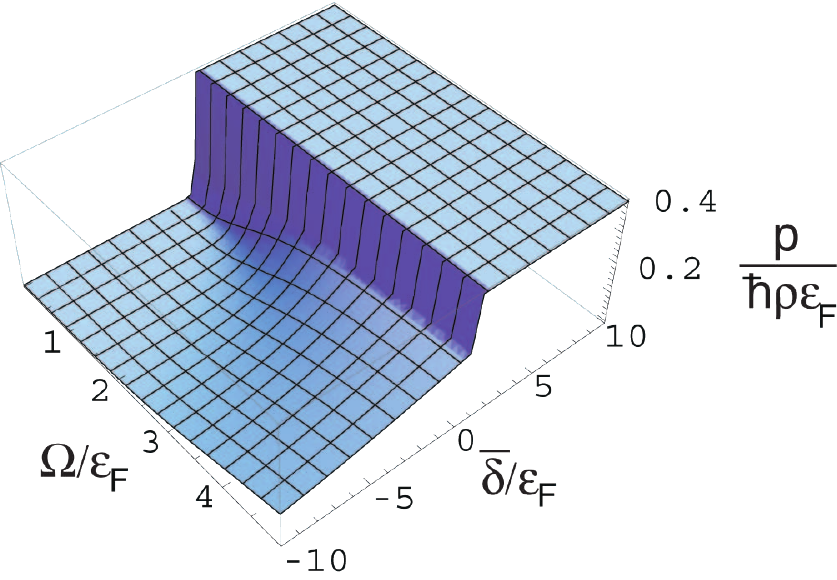}
\end{center}
\caption{(Color online). Pressure $p$ of two-component Fermi gas plotted as a function of
the interaction strength $\Omega$ and detuning $\bar\delta$. The definition
of the interaction strength $\Omega$ is as in Ref.~\cite{JAV05}, but, for
compatibility with the present paper, the detuning here is the detuning of
Ref.~\cite{JAV05} divided by two. For comparison, for an ideal Fermi gas at
zero temperature, $p=\frac{2}{5}\hbar\rho\epsilon_F$.}
\label{FERMIPRESSURE}
\end{figure}

The situation becomes more peculiar after a look at
Fig.~\ref{PRESSURE}, which plots the negative of the pressure for a range
of the parameters
$\Omega$ and
$\bar\delta$. Although the absolute value of the pressure drops sharply toward
of the borderline between the areas of $\alpha=0$ and $\alpha\ne0$ when
entering from the $\alpha\ne0$ side, as far as we can tell, the pressure
remains negative for all values of
$\Omega$ and $\bar\delta$. For instance, the expansions~\eq{EASY}
and~\eq{MUASY} give
\begin{equation}
\frac{p}{\hbar\rho}=-\frac{3 \pi  \Omega ^4}{256 \sqrt{2} \epsilon_F^{3/2}
\left|\bar{\delta }\right|^{3/2}} + {\cal O}(|\bar\delta|^{-2})\,.
\end{equation}
For comparison, we have plotted in Fig.~\ref{FERMIPRESSURE} the pressure of a
two-component Fermi gas from Eq.~\eq{PGD} in complete analogy with
Fig.~\ref{PRESSURE}. No negative pressure develops for a Fermi gas in the
corresponding mean-field theory~\cite{JAV05}. Moreover, there is no question about the
experimental stability of the Fermi gas in the vicinity of the Feshbach
resonance. 

As it comes to the Bose gas, the negative pressure is a mixed blessing. It is an
interesting prediction in its own right, but complicates the observation of the
other features we have discussed. We hope that additional atom-atom,
atom-molecule, and molecule-molecule interactions not included in our model
might stabilize the Bose gas sufficiently, especially on the
$\alpha=0$ side, that the phase transition like feature and the many-body
shift of the Feshbach resonance could be seen under favorable circumstances.

\section{Concluding remarks}

So far the molecular lifetimes have not allowed atom-molecule
equilibration in a Bose gas near a Feshbach resonance, so that the kind of
systems we have studied are presently not feasible  experimentally. However,
progress is continually made~\cite{WIN05} in two-photon (two-color)
photoassociation in a Raman scheme~\cite{WYN00}, and in
heteronuclear systems direct one-photon photoassociation~\cite{OSP06} from
the dissociation continuum to a low-lying vibrational level is possible in
principle. We are of the opinion that photoassociation to a stable molecular
state will eventually be achieved in a BEC.

There is a semantic issue with the mean-field theory, in that, for
instance, the review~\cite{KOH06} regards our approach as going beyond the
mean-field theory. Now, in a usual atomic BEC ``mean-field theory'' means
Gross-Pitaevskii equation for the condensate~\cite{STR99}. The analogy in
atom-molecule systems would, indeed, be an analysis in terms of atomic and
molecular condensates only, with no noncondensate atoms or molecules. This
was the case in the early discussions~\cite{TIM98,JAV99}, which
we nowadays term ``two-mode'' models. 

However, such an analysis ignores an
inherent asymmetry in atom-molecule conversion. According to momentum
conservation, in a Feshbach resonance in free space two zero-momentum
condensate atoms may only combine into a zero-momentum condensate molecule,
but in the reverse process the two atoms dissociated from a zero-momentum
molecule may have arbitrary opposite momenta.
Taking into account such
``rogue''dissociation~\cite{JAV02} was the original aim of our present
modeling that accounts for correlated pairs of atoms. Initially we
regarded this as an extension of mean-field theory, but after we
discovered that the time independent version for fermions is a variation of
the BCS theory, one of the quintessential mean-field theories, we refer to our
approach simply as mean-field theory.

Mean-field theories, of course, are an interesting case in their own right.
They are typically the first, and occasionally the last, tool in the
analysis of a new phase transition, even though it is well known that they
cannot be expected to be quantitatively accurate. In fact, we cannot
think of any instance in the area of quantum degenerate dilute gases in
which a properly formulated mean-field theory has given a qualitatively wrong
prediction, and only very recently has it become possible to
distinguish between  mean-field theory and a strongly correlated approach quantitatively in
an experiment; c.f.\ Refs.~\cite{LUO07} and~\cite{BUL07}.

These observations cast an interesting light on our prediction of negative
pressure even on the molecule side of the Feshbach resonance, where the
standard picture is that the scattering length should be large and positive.
Granted, our calculations are missing the usual background scattering length
and are cursory about the structure and collisions of the molecules.
One may also argue that, as we have no time scale to offer for the associated
instability, it may be unobservable even if it really existed. After all, at
sub-Kelvin temperatures, at any pressure, the thermodynamic ground state of
alkali metals is a solid not a gas, which has not precluded innumerable
successful BEC experiments.

Nonetheless, the negative
pressure may be
viewed as an opportunity rather than a nuisance. It either exists, or is a
qualitatively wrong prediction from a mean-field theory. Speculating further,
since we do not have any time scale for the collapse associated with the
negative pressure, we do not know that it is short either. It is a well-known
empirical fact that in the neighborhood of a Feshbach resonance, on both
sides of the resonance, atoms are lost from a BEC. Perhaps the negative
pressure contributes to, or even dominates, the loss.

We close with a remark about a rather esoteric original motivation for this
particular piece of work on bosons. Some time ago we noticed~\cite{YIN95} in a
simple model for two trapped ions that the statistics has no effect on the
thermodynamic properties. Our interpretation was that the Coulomb interaction
between the ions keeps them sufficiently far apart, so that they are
effectively distinguishable and quantum statistics is moot. Calculations in
which strongly interacting bosons crystallize~\cite{ROM04} in a trap just
like ions do then led us to the idea of ``superuniversality:'' If
strong interactions between the particles keep them apart, the state
could not only not depend on the strength of the interactions, but could also
be independent of atom statistics. Now, within the confines of
mean-field theory, our results for bosons and fermions are quite different.
Technically, superuniversality in mean-field theories is neither a necessary
nor a sufficient condition for superuniversality in Nature, but nonetheless, the
notion of superuniversality did not pass this particular test.
\acknowledgments
This work is supported in part by NSF
(PHY-0354599, PHY-0750668) and NASA (NAG3-2880).
\appendix
\section{Steady state}
\label{FANOSS}
In this Appendix we study the steady state of the two-atom model as a
bound-continuum problem along the lines of the classic paper by Fano~\cite{FAN61}, with
due consideration to renormalization of the ultraviolet divergence.

We imagine that there is initially a bound molecular state
$|b\rangle$ and a continuum of states $|\epsilon\rangle$ labeled by frequency
and normalized in such a way that
$\langle\epsilon|\epsilon'\rangle=\delta(\epsilon-\epsilon')$, coupled by a
photoassociative or Feshbach resonance coupling. We write the Hamiltonian as
\begin{equation}
\frac{H}{\hbar}=|b\rangle\delta\langle b|+\int d\epsilon\,
|\epsilon\rangle\delta\langle \epsilon|+\int d\epsilon\,K(\epsilon)
\left[
|b\rangle\langle \epsilon|+|\epsilon\rangle\langle b|
\label{CONTHAM}
\right]\,,
\end{equation}
with 
\begin{equation}
K(\epsilon)=\sqrt[4]{\frac{4\kappa\epsilon}{\pi^2}}\,.
\label{KDEF}
\end{equation}
The relevant features in Eq.~\eq{KDEF} are the fourth root of frequency (energy), and the
constant $\kappa$ with the dimension of frequency; the rest of the constants
are an attempt to simplify the appearance of a few results below. Similarly, the state vector is
\begin{equation}
|\psi\rangle = b(t)|b\rangle + \int d\epsilon\,a(\epsilon;t)|\epsilon\rangle.
\end{equation}

The time dependent Schr\"odinger equation becomes
\bea
i\frac{\partial}{\partial t}b &=& \delta b + \int d\epsilon\, K(\epsilon)a(\epsilon),\label{FANEQBTD}\\
i\frac{\partial}{\partial t}a(\epsilon) &=&\epsilon\, a(\epsilon) +K(\epsilon)b\,,\label{FANEQATD}
\eea
and the time
independent Schr\"odinger equation for an eigenstate of the
Hamiltonian with the eigenfrequency $\omega$ is
\bea
(\omega-\delta)b &=& \int d\epsilon\, K(\epsilon)a(\epsilon),\label{FANEQB}\\
(\omega-\epsilon)a(\epsilon) &=& K(\epsilon)b\,.\label{FANEQA}
\eea
The substitution
\begin{equation}
a(\epsilon)=-\frac{3 \sqrt{\pi } \sqrt[4]{\epsilon }\, \Omega 
}{\sqrt{2} \sqrt[4]{\kappa }\, \epsilon 
_F^{3/2}}A\left(\frac{\epsilon }{2}\right)
\end{equation}
converts Eqs.~\eq{FANEQB} and~\eq{FANEQA} to Eqs.~\eq{STATBETAEQ}
and~\eq{STATAEQ}. Moreover, let us solve Eqs.~\eq{FANEQBTD} and~\eq{FANEQATD} in
the pole approximation for a positive detuning $\delta$, and likewise solve
Eqs.~\eq{SCHRBETAEQ} and~\eq{SCHRAEQ} in the pole (Wigner-Weisskopf)
approximation, then the ensuing decay
rates for the molecules are equal if the parameters are related by
\begin{equation}
\kappa = \frac{9 \pi^2\Omega^4}{512 \epsilon_F^3}\,.
\label{CONTVSDISCR}
\end{equation}

The present continuum problem~\eq{CONTHAM} simply solves the dilute-gas
(single molecule) limit of our mean-field theory, but with the advantage that
the amplitudes
$b$ and $a(\epsilon)$ also have the normalization conditions that follow from
the orthonormality conditions of the states $|b\rangle$ and $|\epsilon\rangle$
and the manifest hermiticity of Eqs.~\eq{FANEQBTD} and~\eq{FANEQATD}. In the steady state the
amplitudes $b$ and $a(\epsilon)$, of course, depend on the eigenvalue
$\omega$, a dependence that we will write down below.

Now, the formal solution for $a(\epsilon)$ of Eq.~\eq{FANEQA},
\begin{equation}
a(\epsilon,\omega) = \frac{K(\epsilon)b(\omega)}{\omega-\epsilon}\,,
\label{CONTAMP}
\end{equation}
contains a singularity at $\epsilon=\omega$ that renders the meaning of the
right-hand side of Eq.~\eq{FANEQB} ambiguous. The substitution
$\omega\rightarrow\omega-i\eta$, with $\eta=0+$, removes the
ambiguity, but gives  at most one stationary state. Following Fano~\cite{FAN61}, we therefore attempt a  solution in the form
\begin{equation}
a(\epsilon,\omega) =
\left[{\cal
P}\frac{1}{\omega-\epsilon}+f(\omega)\delta(\epsilon-\omega)\right]
K(\epsilon)b(\omega)\,,
\label{ASUBEQ}
\end{equation} 
where $\cal P$ denotes the principal value integral and $f(\omega)$ is yet to
be determined; $f(\omega)=i\pi$ would give the forward-in-time solutions we
have discussed earlier. Inserting Eq.~\eq{ASUBEQ} into Eq.~\eq{FANEQB}
gives
\begin{equation}
\omega-\delta = \theta(\omega)f(\omega)K^2(\omega)
+{\cal P}\int d\epsilon\,\frac{K^2(\epsilon)}{\omega-\epsilon}\,,
\end{equation}
with $\theta$ being the usual unit step function. The integral on the
right-hand side has the same ultraviolet divergence as before, and the cure
is exactly the same; we add the infinity $\int d\epsilon\,
K^2(\epsilon)/\epsilon$ to both sides of the equation, which
renormalizes the detuning and makes the principal value integral convergent,
\begin{equation}
\omega-\bar\delta = \theta(\omega)f(\omega)K^2(\omega)
+{\cal P}\int
d\epsilon\,\frac{K^2(\epsilon)\omega}{\epsilon(\omega-\epsilon)}\,.
\label{RENENE}
\end{equation}

\subsection{Bound State}
Suppose first that $\omega<0$. Then the principal value integral is
a usual integral, and we have the equation
\begin{equation}
\omega-\bar\delta= 
2\sqrt{-\kappa 
\omega }\,.
\end{equation}
It turns out that this equation has a real solution if and only if
$\bar\delta<0$, and the unique solution is then
\begin{equation}
\omega_b=\bar{\delta }-2 \kappa +
2 \sqrt{\kappa ^2-\kappa  \bar{\delta }}\,.
\label{BSENE}
\end{equation}
There is one, and only one, negative-energy
solution if and only if the detuning is negative.

One might surmise that the negative-energy solution is bounded, i.e.,
normalizable to unity. In fact, using Eq.~\eq{CONTAMP}, we find
\begin{equation}
\int d\epsilon\,|a(\epsilon,\omega_b)|^2 = \sqrt{\frac{\kappa }{-\omega
_b}}\,\,|b(\omega_b)|^2\,,
\end{equation}
so that the bound and continuum amplitudes in the wave function normalized to
unity are
\bea
b(\omega_b) &=&
\sqrt{\frac{\sqrt{-\omega_b}}{\sqrt{\kappa}+\sqrt{-\omega_b}}}\,,\\
a(\epsilon,\omega_b)&=&\frac{\sqrt[4]{\kappa\epsilon/\pi^2
}}{\omega_b-\epsilon}\,b(\omega_b)\,.
\eea

\subsection{Continuum States}
The system also has a bountiful of positive-energy eigenstates. It
turns out that for $\omega>0$
\begin{equation}
{\cal P}\int
d\epsilon\,\frac{K^2(\epsilon)\omega}{\epsilon(\omega-\epsilon)}=0\,,
\label{PRIZERO}
\end{equation}
so that Eq.~\eq{RENENE} gives
\begin{equation}
f(\omega) = \frac{\omega-\bar\delta}{K^2(\omega)}\,.
\end{equation}
For positive energies we are obviously dealing with continuum states, so that
we aim at the normalization
\begin{equation}
b^*(\omega) b(\omega') + \int
d\epsilon\,a^*(\epsilon,\omega)a(\epsilon,\omega')=\delta(\omega-\omega')\,.
\end{equation}
Note that while for a unit-normalized bound state the coefficient
$b(\omega_g)$ is dimensionless, here the dimension of $b(\omega)$ will be
the inverse of the square root of frequency. The calculation  based  on
Eq.~\eq{ASUBEQ} is straightforward except for the ensuing product of
principal-value integrals, which may be handled as in~\cite{FAN61};
\bea
&&{\cal P}\frac{1}{\omega-\epsilon}\,{\cal P}\frac{1}{\omega'-\epsilon}
=\frac{1}{\omega'-\omega}\left(
{\cal P}\frac{1}{\omega-\epsilon}-{\cal
P}\frac{1}{\omega'-\epsilon}
\right)\nonumber\\
&&\qquad + \pi^2\delta(\omega-\omega')\delta(\epsilon-\omega)\,.
\eea
The principal value integrals arising in this way give zero just as in
Eq.~\eq{PRIZERO}, so we have
\begin{equation}
|b(\omega)|^2\left[\pi^2
K^2(\omega)+\frac{(\omega-\bar\delta)^2}{K^2(\omega)}
\right]=1\,.
\end{equation}
\begin{widetext}
The continuum state vectors for $\omega>0$ are therefore fully specified by
\bea
b(\omega) = \sqrt{\frac{2}{\pi }}\frac{ \sqrt[4]{\kappa 
\omega}}{\sqrt{\left(\omega -\bar{\delta }\right)^2+4 \kappa 
   \omega }}\,;\qquad a(\epsilon,\omega) 
=\left[\sqrt[4]{\frac{4 \kappa  \omega}{\pi^2} }{\cal
P}\frac{1}{\omega-\epsilon}+(\omega-\bar\delta)^2
\sqrt[4]{\frac{\pi^2}{4 \kappa  \omega}
}\delta(\epsilon-\omega)\right]b(\omega)\,.
\eea
\end{widetext}

\end{document}